\documentclass[amsmath,amssymb,prc,twocolumn,runinadress,superscriptaddress,showpacs,nofootinbib,floatfix]{revtex4-1}
\pdfoutput=1
\usepackage{graphicx}
\usepackage{epsf}
\usepackage{textcomp,color}
\usepackage{amsmath}
\usepackage{natbib}
\usepackage{dcolumn}
\usepackage{bm}

\def\topfraction{1.0}
\renewcommand{\topfraction}{1.0}

\newcommand{\beq}{\begin{equation}}
\newcommand{\eeq}{\end{equation}}
\newcommand{\beqa}{\begin{eqnarray}}
\newcommand{\eeqa}{\end{eqnarray}}

\def\2nudbd{$2 \nu \beta \beta$ decay}

\def\jpi{$J^{\pi}$}

\def\znbb{$0\nu \beta \beta$}
\def\bb{$\beta \beta$}
\def\bmbm{$\beta^- \beta^-$}

\def\se76{$^{76}$Se}
\def\as76{$^{76}$As}
\def\ti48{$^{48}$Ti}
\def\kr82{$^{82}$Kr}
\def\ba133{$^{133}$Ba}
\def\eu152{$^{152}$Eu}
\def\co60{$^{60}$Co}
\def\am241{$^{241}$Am}

\def\cd106{$^{106}$Cd}
 \def\ag106{$^{106}$Ag}
 \def\sm150{$^{150}$Sm}

\def\gray{$\gamma$-ray}

\def\mucap{$\mu$-capture}
\def\muXray{$\mu$X-ray}
\def\gam{$\gamma$}
\newcommand{\dubna}{\affiliation{Joint Institute for Nuclear Research, 141980 Dubna, Russia}}
\newcommand{\dubnaMUD}{\affiliation{State University ``Dubna'', 141980 Dubna, Russia}}
\newcommand{\PSI}{\affiliation{Paul Scherrer Institute, 5232 Villigen, Switzerland}}
\newcommand{\finland}{\affiliation{Department of Physics, University of  Jyv\"{a}skyl\"{a}, PO Box 35, FIN-40351  Jyv\"{a}skyl\"{a}, Finland}}

\begin{document}

\preprint{To be submitted to Physical Review C}

\title{Ordinary muon capture studies for the matrix elements in $\beta\beta$ decay}

\author{D. Zinatulina}  \dubna
\author{V. Brudanin}    \dubna
\author{V. Egorov}      \dubna\dubnaMUD
\author{C. Petitjean}   \PSI
\author{M. Shirchenko}  \dubna
\author{J. Suhonen}  \finland
\author{I. Yutlandov}   \dubna

\date{\today}

\begin{abstract}
{HPGe detectors were used to make a precise measurement of the $\gamma$-ray spectrum produced following ordinary (non-radiative) capture of negative muons by natural Se, Kr, Cd and Sm. The measurement was repeated for  isotopically-enriched $^{48}$Ti, $^{76}$Se, $^{82}$Kr, $^{106}$Cd and  $^{150}$Sm targets. By investigating energy and time distributions, the lifetime of negative muons in the different isotopes was deduced.
A detailed analysis of the intensity of the $\gamma$-lines enabled the extraction of the relative yields of several daughter nuclei. The partial rates of $(\mu^-,\nu)$ capture to numerous excited levels of the $^{48}$Sc, $^{76}$As, $^{82}$Br, $^{106}$Ag and $^{150}$Tc isotopes (considered to be virtual states of the intermediate odd-odd nuclei in the \bb\ decay of $^{48}$Ca, $^{76}$Ge, $^{82}$Se, $^{106}$Cd and $^{150}$Nd, respectively) were also extracted. These rates are important as an experimental input for the theoretical calculation of the nuclear matrix elements in \bb\ decay.}
\end{abstract}


\maketitle

\renewcommand{\topfraction}{1.0}

\section{INTRODUCTION} \label{sec:introduction}

The search for the hypothetical neutrinoless double-beta \znbb~decay of atomic nuclei is currently the only practical method of determining the Majorana nature of the neutrino.
In order to occur, the decay requires the violation of lepton-number conservation, and a non-zero neutrino mass.
Due to the importance of the related beyond-the-Standard-Model physics,
it is of interest to study \bb-decaying nuclei, both by experimental and theoretical means. Large
experimental collaborations have been established in order to measure
the $0\nu\beta\beta$ half-lives in current and future
underground experiments. The connection between any measured half-life and the fundamental observables, like the electron neutrino
mass, is provided by the nuclear matrix elements (NMEs)~\cite{Suh1998}.

\begin{figure}[h]
 \begin{center}
   \includegraphics [width=\linewidth]{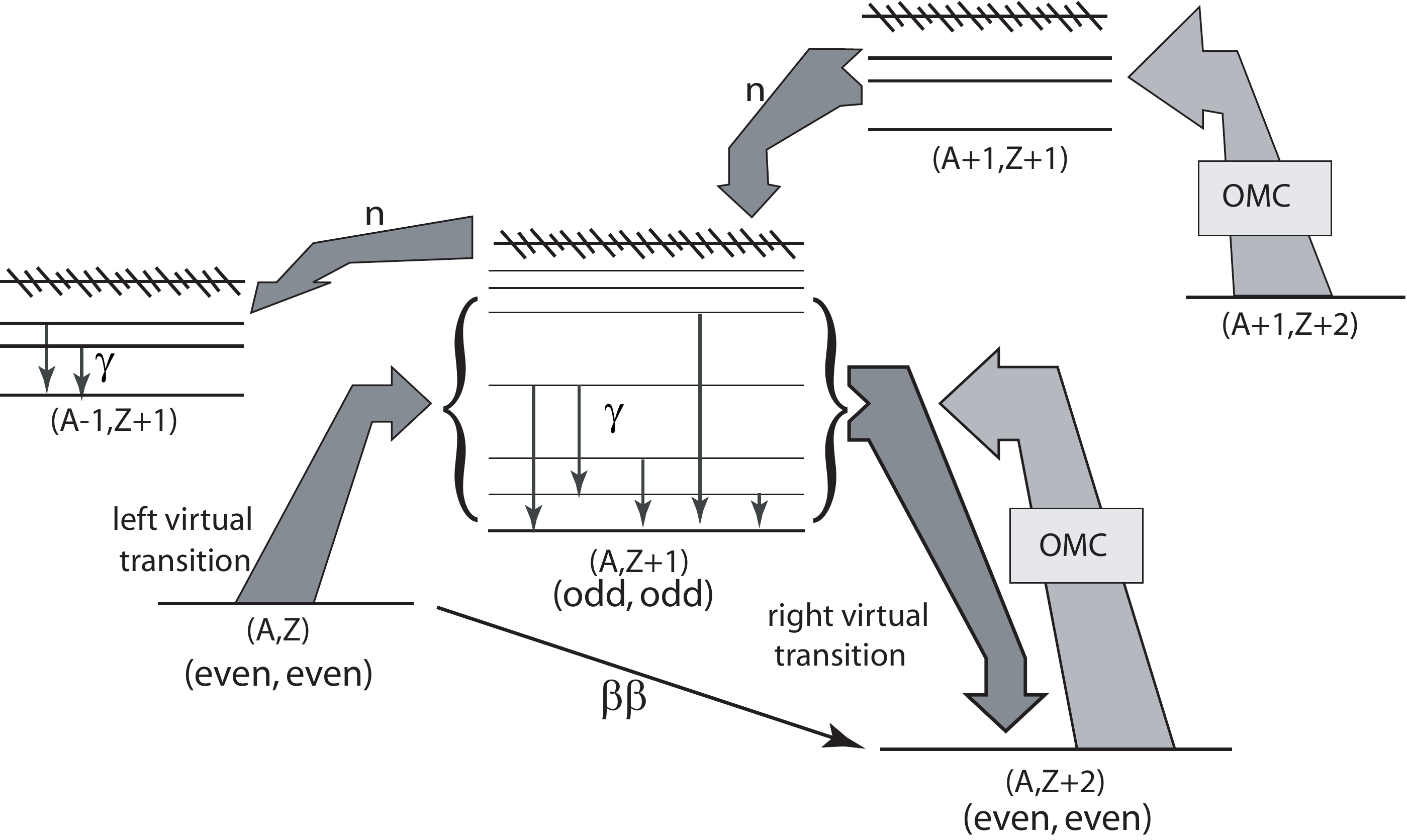}
 \caption{Schematic representation of \bb\ decay of an (even, even) nucleus as two consequent virtual transitions via excited states of an intermediate (odd, odd) nucleus. OMC on a target consisting of the (A,Z+2) daughter nucleus could provide information about the right virtual transition. Any admixture of
the heavier (A+1,Z+2) isotope would cause intensive population of the same intermediate states, and would thus be a background to the measurement.}
 \label{fig:sketch}
 \end{center}
\end{figure}

Nuclear models aimed at the description of the NMEs of $0\nu\beta\beta$
decays have traditionally been tested in connection with
two-neutrino $\beta\beta$ ($2\nu\beta\beta$) decays~\cite{Suh1998,Eng2017} and $\beta$ decays~\cite{Suh2005}.
In~\cite{Kor2002} it was proposed that the ordinary muon capture (OMC)
could also be used for this purpose, as well. The $2\nu\beta\beta$ and $\beta$
decays are low-momentum-exchange processes ($q\sim$ a few MeV), whereas both
$0\nu\beta\beta$ and OMC are high-momentum-exchange processes ($q\sim 100$ MeV).
In this way the $0\nu\beta\beta$ and OMC are similar processes and possess
similar features: they are able to excite high-lying nuclear states with
multipolarities $J^{\pi}$ higher than $J^{\pi}=1^+$. The $0\nu\beta\beta$
decay proceeds between the $0^+$ ground states of parent and daughter
even-even nuclei through virtual states of the intermediate odd-odd nucleus.
These same virtual states can be accessed by the OMC from either the
daughter nucleus (electron-emitting $\beta\beta$ decays, see Fig.~\ref{fig:sketch}) or
the parent nucleus (positron-emitting/electron-capture $\beta\beta$ decays).

The ability of the OMC to access high-energy and high-multipolarity virtual
states of the $0\nu\beta\beta$ decay is a strong asset of the process, as is the similarity of the two processes in terms of the energy scale
of the momentum exchange. There are also differences between the processes,
stemming from factors such as the neutrino potential generated by the propagator
of the virtual Majorana neutrino in the $0\nu\beta\beta$ decay~\cite{Doi1985}.
Despite this difference, the OMC can effectively probe the nuclear wave
functions relevant for the $0\nu\beta\beta$ decay, as shown for the light
nuclei in the shell-model framework in~\cite{Kor2004}.

For the medium-heavy and heavy open-shell nuclei, the shell-model framework
is unfeasible due to computational limitations. For these nuclei, the
model framework of the quasiparticle random-phase approximation
(QRPA)~\cite{Suh2007} is a good choice. In particular, the proton-neutron
version of the QRPA (pnQRPA) can access the virtual intermediate states of the
$0\nu\beta\beta$ decays~\cite{Suh1998}. The key issue with the
pnQRPA approach is the uncertainty associated with one of its key parameters,
the particle-particle interaction strength $g_{\rm pp}$. This parameter
is used to introduce a phenomenological overall scaling of the
particle-particle part of the proton-neutron interaction~\cite{Vog1986}. It is not
clear how this scaling should be modelled for $0\nu\beta\beta$ decays since
there is no experimental data for transitions from either the $0\nu\beta\beta$
mother or daughter nuclei to the multipole $J^{\pi}\ne 1^+,2^-$ intermediate
states. The $1^+$, and to some extent, the $2^-$, states can be probed by the $(p,n)$ and
$(n,p)$ charge-exchange reactions~\cite{Fre2017}. In this case the only
viable method to investigate this ``$g_{\rm pp}$ problem'' is the OMC~\cite{Kor2003}.
By using experimental data on OMC to individual intermediate $J^{\pi}$ states,
one can access the value of $g_{\rm pp}$ for each multipole separately, and at the same time can
study the consistency of these values, by comparing with the measured OMC rates for a wider palette of nuclear states.

In order to give an experimental input to \bb\ NME calculation, it was proposed in~\cite{Kor2002,R0202} to measure partial muon capture rates in corresponding isotopes (Fig.~\ref{fig:sketch}). Due to a high momentum transfer, numerous excited states of the intermediate $(A,Z+1)$ nucleus are populated, and thus their ``functions can be touched'' through the transition intensities.

It should be mentioned that the OMC reaction $A(\mu^-,\nu)$
is typically an order of magnitude less probable than the similar process where the muon capture is followed by the emission of one or two neutrons: $(\mu^-,\nu~n)$, $(\mu^-,\nu~2n)$. As a result, even a small contamination of the target with heavier isotopes $(A+1)$ or $(A+2)$ could cause significant background, as $(\mu^-,\nu~n)$, $(\mu^-,\nu~2n)$ reactions in the contaminating isotope would populate the same excited daughter states as OMC interactions in the target.
It is therefore very important to use isotopically enriched targets without significant content of $(A+1)$ or $(A+2)$ isotopes.
In this work we describe \mucap\ measurements on isotopically enriched  \ti48,  \se76, \kr82,  and \sm150, all products of \bmbm\ decay, and on \cd106, which is the parent nucleus in $(\beta^+\beta^+)/(\beta^+{\rm EC)/(ECEC)}$ decay. The preliminary results of this work are also published in~\cite{Zinatulina-AIP1894-2017}.

\section{EXPERIMENT}

\subsection{Different types of events registered in the OMC experiment.}\label{subsec:UPD}

All information from $\gamma$ detectors can be divided into two types -- events correlated and uncorrelated in time with incoming muons.
If a signal from the $\gamma$ detector was not preceded by a muon stopped in the target within some time window $W$ \footnote{The duration of $W$ should be set at a level of few microsecond depending on the expected muon life time in the given target.} -- then the event is considered as {\sl Uncorrelated}.
To save disk space, uncorrelated events may not be individually recorded, but their energies will be used to generate a U-spectrum of uncorrelated energy measurements.

\begin{figure}[h]
 \begin{center}
   \includegraphics [width=\linewidth]{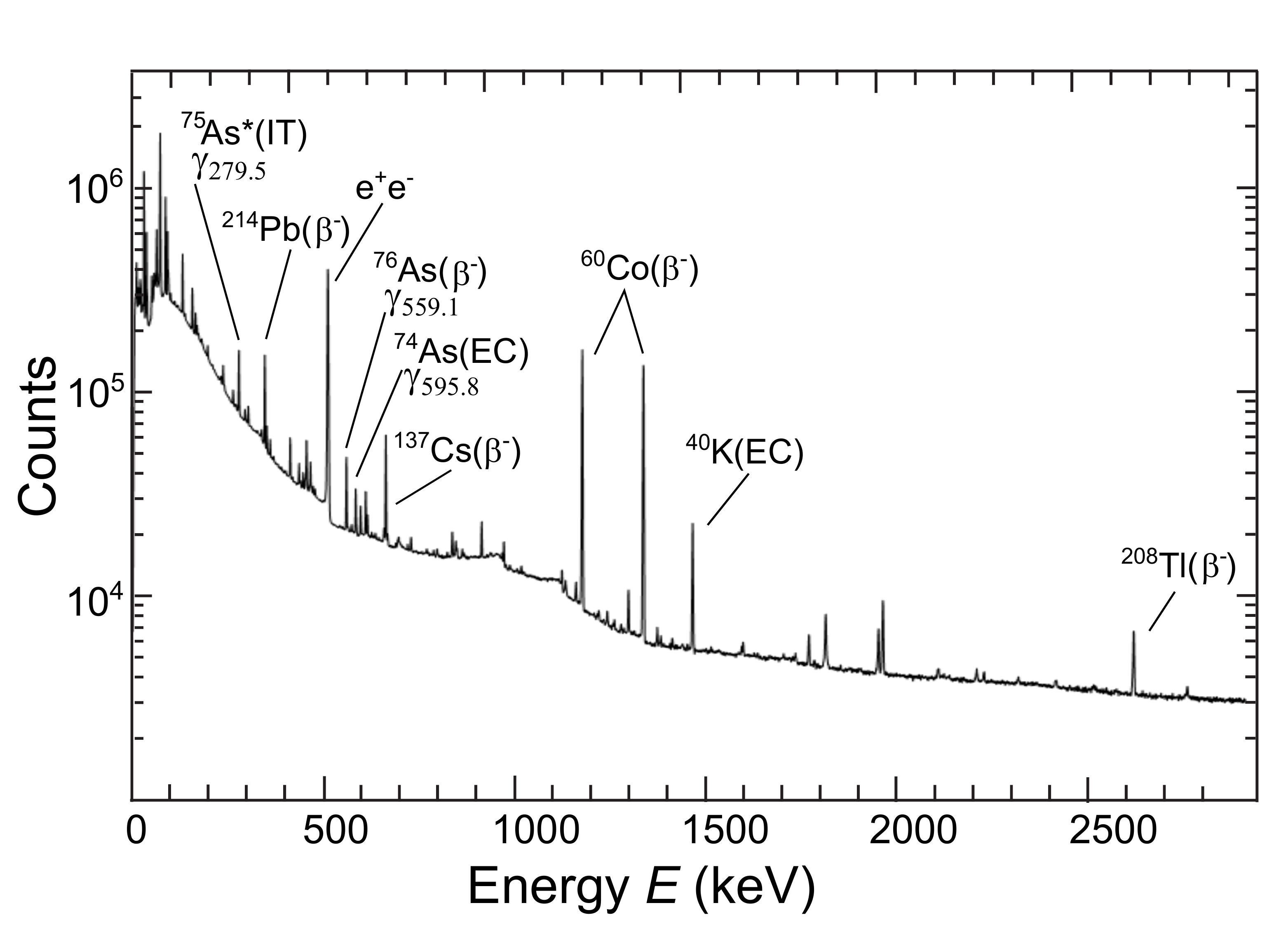}
 \caption{An example of U-spectrum measured with a HPGe detector and the $^{76}$Se target. Some of $\gamma$-lines follow decay of the OMC products -- $^{74m}$As,  $^{75}$As, $^{76}$As.}
 \label{fig:Uspectra_76Se}
 \end{center}
\end{figure}

A typical U-spectrum (Fig.~\ref{fig:Uspectra_76Se}) includes $\gamma$-lines of natural ($^{40}$K, U- and Th-chains) and man-made ($^{60}$Co, $^{137}$Cs) background, as well as beam-induced $(n,\gamma)$-reactions. The above lines could be used to calibrate the detectors. The OMC products are frequently unstable, emitting $\gamma$-rays as they decay. The intensity of these $\gamma$-rays in the U-spectrum allows the yield of the individual isotopes and isomers to be extracted.

\begin{figure}[h]
 \begin{center}
   \includegraphics [width=\linewidth]{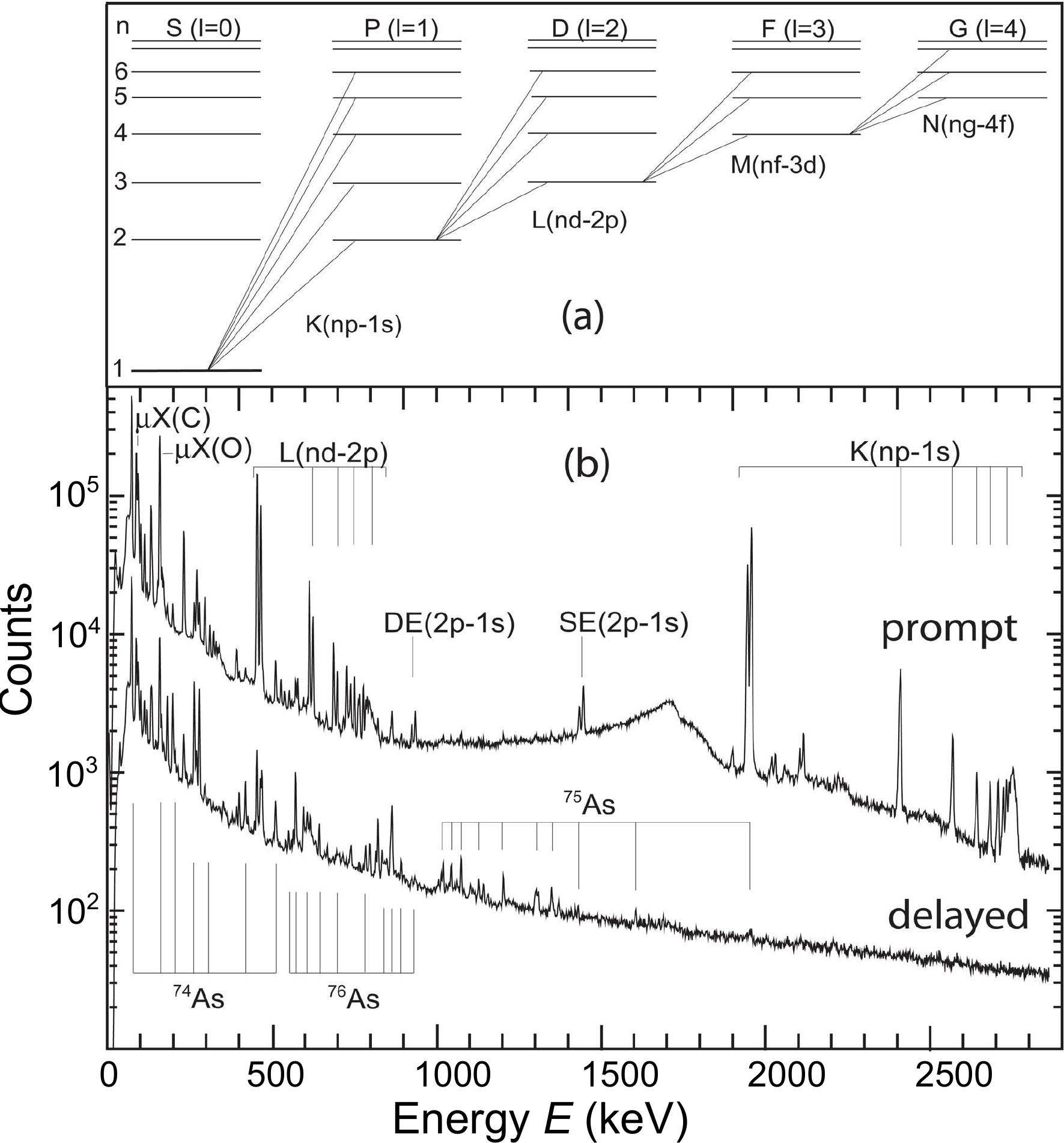}
 \caption{(a) -- a simplified scheme of the transitions in a muonic atom. (b) -- correlated P- and D-spectra measured with enriched $^{76}$Se target.
 Some of the identified \gray\ transitions are indicated.``DE" and ``SE" denote the double and single escape peaks. The \muXray\ transitions for carbon and oxygen are indicated as $\mu$X(C) and $\mu$X(O).}
 \label{fig:Se-prompt-delayed}
 \end{center}
\end{figure}

{\sl Correlated} events - those which occur within the time
window $W$ immediately after the $\mu$-stop - are more informative. The majority of these events are caused by a cascade of so-called muonic X-rays ($\mu$X) -- high energy photons emitted by muonic atom during its transition to the 1s-state from a Rydberg state (Fig.~\ref{fig:Se-prompt-delayed}a). The de-excitation process takes place within picoseconds - negligible for this measurement - meaning that \muXray can be considered {\sl Prompt} radiation.

\begin{figure}[h]
 \begin{center}
   \includegraphics [width=\linewidth]{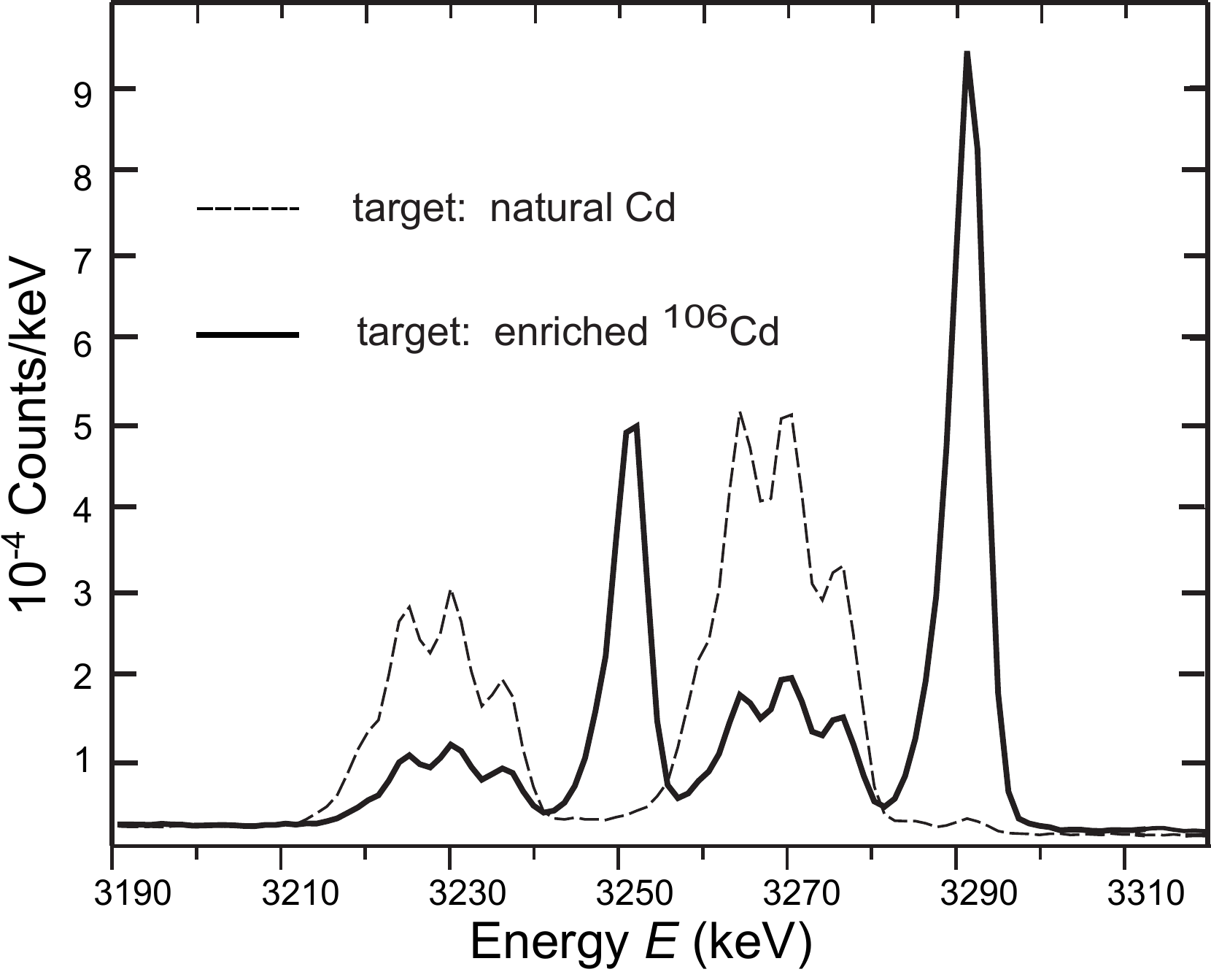}
 \caption{$K\alpha_1$ and $K\alpha_2$ $\mu$X-ray doublet measured with enriched and natural Cd targets.}
 \label{fig:isotope-shift}
 \end{center}
\end{figure}

A muon is heavier than an electron by a factor of $\sim$200. Therefore, energy of $\mu$X is higher by the same factor with respect to conventional X-rays, and the radius of the muonic orbit is comparable with nuclear dimensions. In fact, for muonic atoms with $Z\gtrsim 40 \ldots 50$, the muon's orbit is within the nucleus. As a result, an energy spectrum of $\mu$X-rays becomes very sensitive to the collective properties of the nucleus (mass and charge distribution, etc.). Fig.~\ref{fig:isotope-shift} demonstrates isotopic shift of $2p_{3/2,1/2}\rightarrow 1s_{1/2}$ transition doublet measured with two targets: isotopically enriched (enrichment of $^{106}$Cd is about 63\%) and natural (abundance of $^{106}$Cd is 50 times lower -- 1.25\%). A dedicated investigation of muonic X-rays is outside the scope of this work. However, as a by-product of this study, we have generated a catalogue of our muonic X-ray measurements~\cite{zinatulina-EPJ-catalogue}; it is available in interactive mode at~\cite{MXcatalog}.

As each $\mu$-stop in a target is followed by characteristic $\mu$X-rays, the intensity of each spectral line reflects the number of $\mu$-stops in the corresponding isotope. Therefore $\mu$X spectra could be used to normalize any measurement by the number of muons.

In addition to these prompt events, {\sl Delayed} correlated events correspond to the nuclear $\gamma$ radiation following
muon capture in the $(\mu^-,\nu~xn)$ reactions. This particular radiation is the main subject of our work.

\begin{figure}[h]
 \begin{center}
   \includegraphics [width=\linewidth]{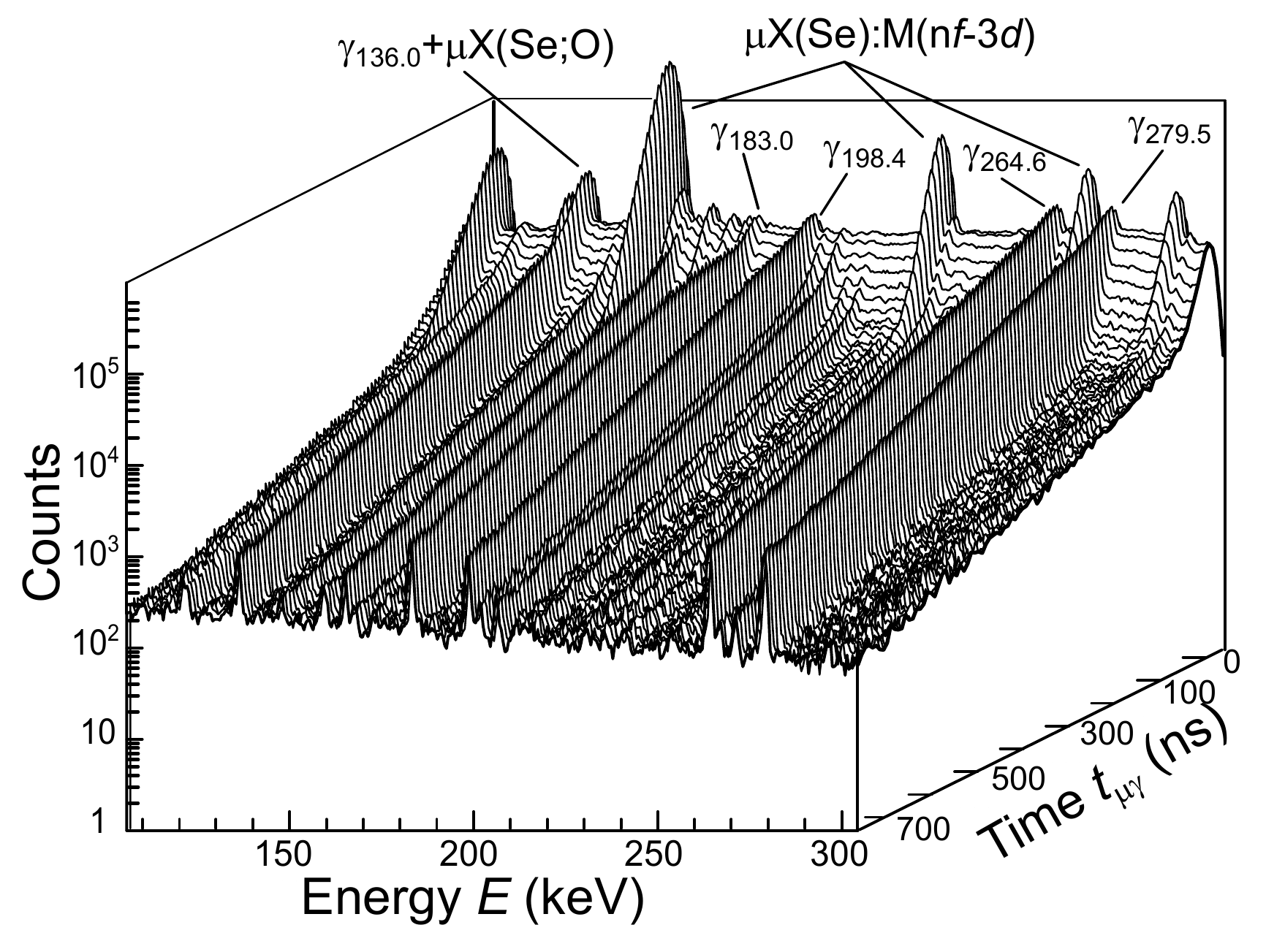}
 \caption{(E,t) distribution of the correlated events measured with the $^{76}$Se target.}
 \label{fig:Matrix_76Se}
 \end{center}
\end{figure}

As it is related to muon decay, the rate at which the OMC process takes
place has a probability distribution. As a result, the corresponding $\gamma$-rays do not appear immediately, but have an exponentially-distributed delay (Fig.~\ref{fig:Matrix_76Se}). The exponent depends on the OMC probability.

By analyzing the time evolution of the individual $\gamma$-line(s), one can
extract the OMC probability for each corresponding product (contrast with the commonly used inclusive method, which is based on the detection of $\mu$-decay electrons~\cite{Measday-PR354-2001}). The relative intensity of
the delayed lines (with respect to $\mu$X-rays) allows the extraction of partial capture rates to the individual excited states of the daughter nucleus. Prompt and Delayed spectra (Fig.~\ref{fig:Se-prompt-delayed}b) could be partially separated from each other using different time cuts (for instance, to simplify $\gamma$-line identification).

\subsection{Experimental approach.}\label{subsec:approach}

In order to optimize the experiment, the following conditions should be satisfied:
\begin{itemize}
\item the muon energy must be high enough that it will enable muons to pass through materials upstream of the target (output window of the beam-line, muon counters, target entrance window, etc.);
\item the beam energy, and its dispersion, must be small enough to ensure that all muons will stop inside the target (especially in the case of gas, or low-mass enriched targets);
\item beam intensity must be high enough to provide good statistics;
\item the intensity must be low enough to prevent random coincidences;
\item the target must be thick enough to stop all the muons;
\item the target must be thin enough to prevent absorption of low-energy $\gamma$-rays;
\item the muon counters must be thick enough to produce good light signal;
\item the muon counters must be thin enough to prevent muons stopping in the counters.
\end{itemize}
Due to these conflicting requirements, only a few beams are able to satisfy all of the above conditions. It was eventually decided that three sets of measurements would be taken at the $\mu$E4 and the $\mu$E1 beam-lines of the Paul Scherrer Institute (Villigen, CH). The beam consisted of negative muons, emitted in a backward direction from pion decay in flight, yielding a muon momentum centered around \mbox{20--30~MeV/$c$} with $\Delta$p/p at the level of 2\%, and an intensity of approximately $10^{4}$ muons per second.

\begin{figure}[h]
 \begin{center}
   \includegraphics [width=\linewidth]{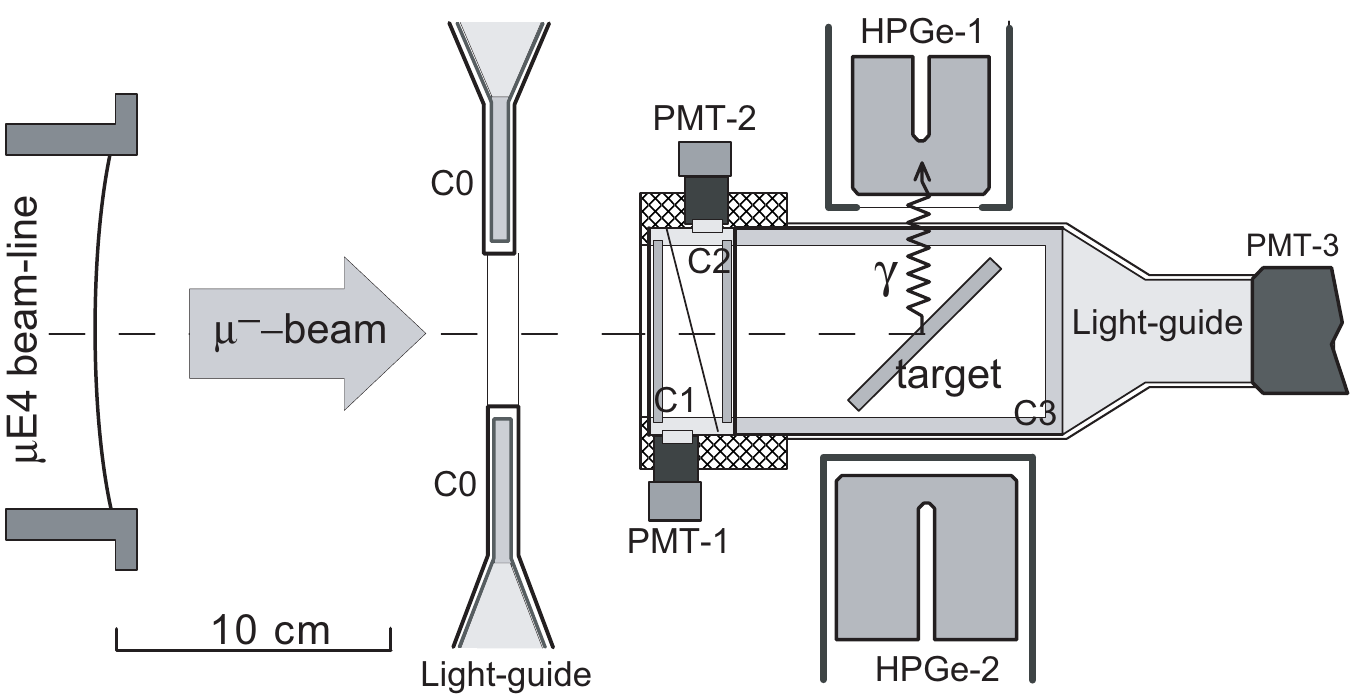}
 \caption{Schematic layout of the $\mu$ beam line and the target arrangement: the aperture defining veto-counters (C0), the trigger counters (C1,C2), the target area, the veto counter (C3) and the HPGe $\gamma$-detectors
 around the target area. The target enclosure was constructed to accommodate solid  (like in this figure) as well as gaseous materials.}
 \label{fig:target}
 \end{center}
\end{figure}

The target arrangement sketched in Fig.~\ref{fig:target} consists of an active muon veto counter system C0 at the entrance of the target enclosure; two  thin (0.5 mm)  pass-through counters, C1 and C2; then the actual target area, surrounded by a cup-like counter C3. The target enclosure is constructed to accommodate both solid and gaseous materials. The C3 counter
plays several roles: being a gas vessel itself, it is used, together with the pass-through counters, to define a $\mu$-stop trigger:
\begin{equation}\label{eq:stops}
\mu_{\rm stop}=\overline{\rm C0}\wedge{\rm C1}\wedge{\rm C2}\wedge\overline{\rm C3},
\end{equation}
as well as to discriminate against high-energy electrons  from  muon decay\footnote{Muon decay in the target area is a background process alternative to OMC. Electrons emitted have very high energy (up to 50~MeV), disturb operation of Ge detectors and produce intensive bremsstrahlung.}. As it is made of low-Z material, C3 does not affect the measurement of low energy \gray s with external detectors.

The beam momentum and position were tuned to maximize the intensity of \muXray~from the target, while minimizing the background from the surrounding material. Under optimal conditions more than 95\% of the data sample corresponded to muons stopped in the target. Typical $\mu_{\rm stop}$ rates during the experiments were between $3\cdot10^3\rm
~s^{-1}$ and $2.5\cdot10^4\rm ~s^{-1}$ and the total acquisition time was between $50\rm~h$ and $150\rm~h$, depending on the target and isotopic enrichment.

The \muXray s and the delayed $\gamma$-rays following  $\mu$-capture were detected with four independent  semi-coaxial HPGe detectors placed around the target at a distance of about 7--10~cm.
The high-energy radiation that accompanies muon decay causes partial saturation of the analog electronics, generating artifacts in the energy spectrum. To reduce
this effect, the shaping time of the spectroscopic amplifiers was kept as short as 1--2~$\mu$s, which worsened the energy resolution, but optimized the general detector response function.

Three large-volume detectors (Ge-1 -- Ge-3) were selected for detection of photon energies up to 10 MeV with good efficiency. The fourth (smaller) detector was dedicated to the detection of low-energy photons, typically below 100 keV, and was therefore equipped with a thin, low-absorptive beryllium entrance window.

In addition to calibrating with standard radioactive sources, several alternative methods were used to calibrate the energy and efficiency\footnote{Here, and henceforth, the ``efficiency'' refers to the probability that the whole of the absorption peak will be contained within the spectrum. The absolute efficiency value would also include the effect of the solid angle.} over the full energy range (up to 10 MeV). The large-volume detectors were calibrated off-line using neutron-induced $\gamma$-rays from a NaCl radiator placed near a $^{252}$Cf neutron source.  The $^{35}$Cl(n,$\gamma$)-reaction provides an intensive high-energy \mbox{$\gamma$-ray} line at 6111.0~keV and  useful lines at
6619.7, 6627.9 and 6642.0~keV. In-beam  neutron-capture processes on $^{56}$Fe and $^{28}$Si produce another series of high-energy \mbox{\gray s}  at well-known relative intensities up to about 8 MeV~\cite{nndc,Measday-PR354-2001}. Prompt
\muXray s from oxygen, titanium, selenium, cadmium, samarium, gold and thorium, with energies ranging from 39 to 6319~keV, were also measured in several ad-hoc runs, each lasting a few hours. Other beam-induced radioactivity is also a source of useful high-energy \gray s (i.e., 6128.6~keV from $^{16}$N decay). To get a wider dynamic range of the energy signals, each of the three large-volume  detectors was connected to two independent spectrometric channels A and B with different gains, thereby providing seven energy signals in total
(cf. Table~\ref{table:detectors}).

\begin{figure}[h]
 \begin{center}
   \includegraphics [width=\linewidth]{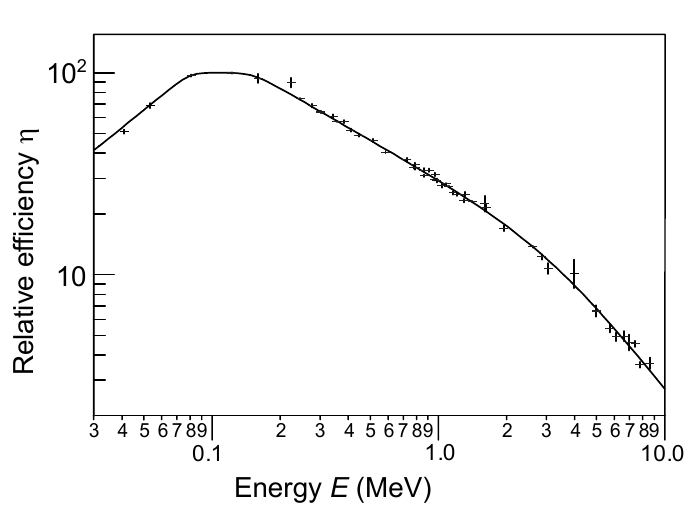}
 \caption{Typical relative efficiency ($\eta$) of the $\gamma$-detection system. In the low energy region (0.03-0.08~MeV) the efficiency
drops  because of absorption in various dead layers. The energy region (0.17-3.2~MeV) is dominated by Compton
scattering, whereas above 3-4~MeV the radiation length gets to be larger than the crystal dimensions.}

 \label{fig:efficiencies}
 \end{center}
\end{figure}

A typical relative efficiency function constructed for the entire photon-energy range is shown in Fig.~\ref{fig:efficiencies}, together with an arbitrary smooth function describing its general energy dependence. The overall error for this function is estimated to be less then 5\%. In the case of \muXray s, the relative efficiency could be normalized by the overall number of $\mu$-stop signals. Absolute efficiency obtained in this way, as well as some of the performance parameters of the various detectors, are quoted in Table~\ref{table:detectors}.

\renewcommand{\arraystretch}{1.2}
\begin{table}[h]
  \begin{center}
    \caption{\label{table:detectors} Performance of the HPGe detectors used in the present OMC experiment.
    The resolution (FWHM) is given at the 1461~keV of $^{40}$K (channel A) and at the 2615~keV $\gamma$-line of $^{232}$Th (channel  B). The energy ranges ($E_{\rm max}$) for both channels are given in MeV.}
   \begin{tabular}{ccccccccc} \hline
    Det. &
    efficiency &  efficiency && \multicolumn{2}{c}{channel  A} & &
    \multicolumn{2}{c}{channel  B}  \\ \cline{5-6} \cline{8-9}
       & @100 keV & @1.0 MeV & &  FWHM &  $E_{\rm max}$ & &  \multicolumn{1}{c}{FWHM} &
    $E_{\rm max}$ \\ \cline{1-3} \cline{5-6} \cline{8-9}

    Ge-1 & 1.32\% & 0.43\% & & 4.34 & 2.85 & &  7.66 & 10.0  \\
    Ge-2 & 1.35\% & 0.39\% & & 3.73 & 2.85 & &  4.81 & 10.0  \\
    Ge-3 & 1.12\% & 0.32\% & & 3.84 & 2.85 & & 10.32 & 10.0  \\
    Ge-4 & 0.65\% & 0.09\% & & 3.58 & 1.60 & &   --- &  ---  \\ \hline
    \end{tabular}
  \end{center}
\end{table}

\begin{figure}[h]
 \begin{center}
   \includegraphics [width=\linewidth]{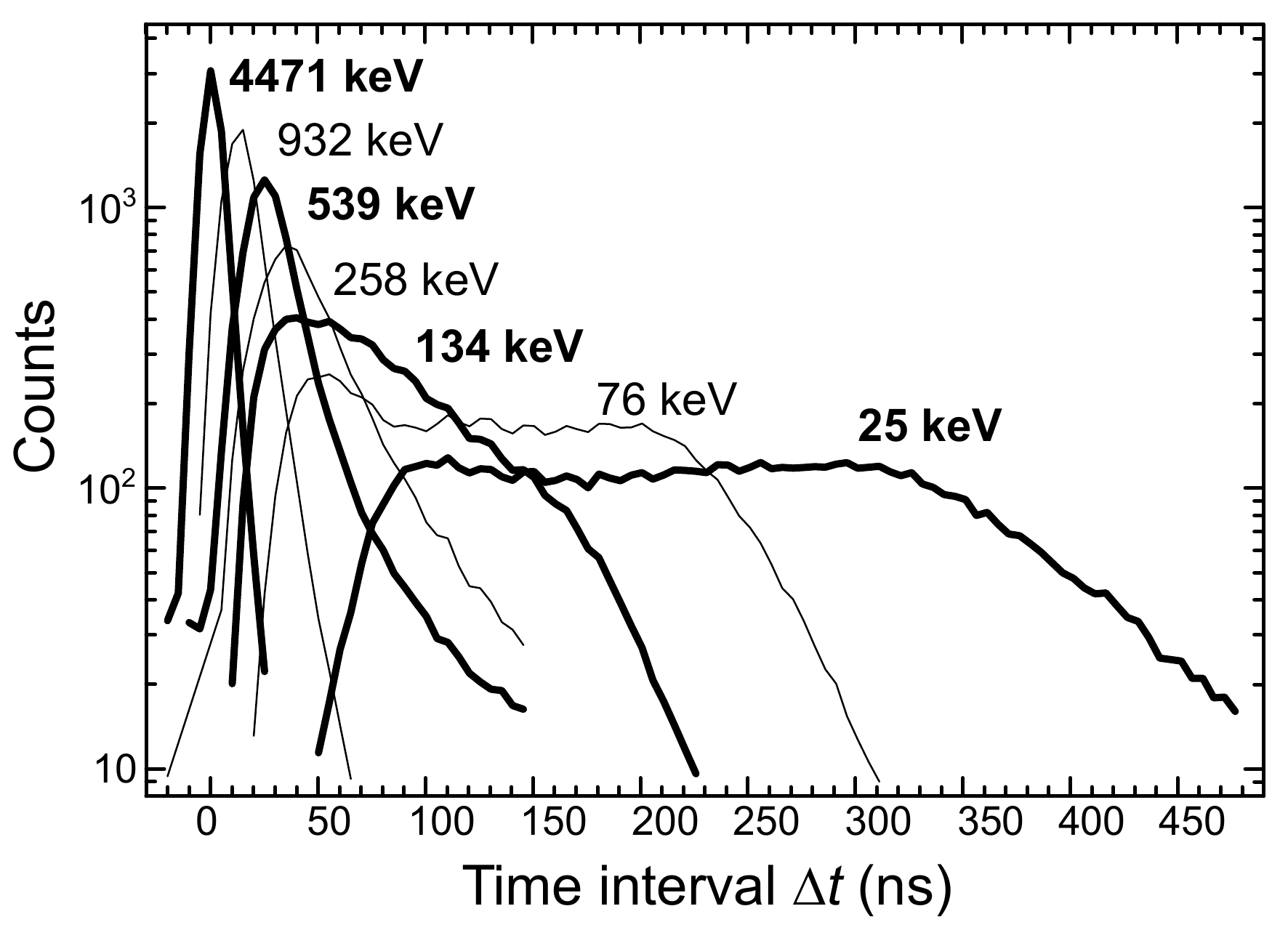}
 \caption{Distribution of the instrumental delay $\Delta$t between the $\mu$-stop and registration of prompt $\mu$X-rays of different energies measured with a big HPGe detector. Time resolution below 200 keV deteriorates due to slow charge collection near the surface edges of the Ge crystal.}
 \label{fig:time-resolution}
 \end{center}
\end{figure}

Large-volume semi-coaxial HPGe detectors are known to have relatively poor timing properties, especially at low \gray\ energy.
Whereas high-energy \gray s are detected more or less homogenously throughout the Ge crystal, low-energy
\gray s are absorbed in a thin layer close to the surface, which includes edge regions with relatively weak and non-uniform electric field. Charge collection in these regions is partially incomplete and spread out in time. As a result, a significant part of the low-energy $\gamma$ signals experiences a time lag of up to a few hundred nanoseconds, and some missing charge, equivalent to up to a few hundred eV. Planar detectors with uniform field are free of this annoying property, but they come at a price of reduced efficiency in the MeV region.
The measured timing characteristics of our current large-volume HPGe detectors are shown in Fig.~\ref{fig:time-resolution}, indicating the rapid deterioration of the timing resolution below about 200~keV.
On the contrary, high energy ~\muXray s are well-suited for monitoring timing stability throughout the experiment.

As both the Correlated and Uncorrelated spectra were
important to us, signals from the HPGe detectors were given
the highest priority, and served as a hard trigger. The event was identified as correlated or uncorrelated, using the $\mu$-stop signal, delayed by the {\sl W} value.

All the measured time and energy information for each target was saved as a separate subdirectory (RUN) once per hour. It consists of files with Uncorrelated (U) energy spectra (4 Ge detectors, A and B channels) and files with Correlated events. Each event includes values of energy and time corresponding to the signals registered by all HPGe detectors, as well as ``flag'' bits from the input register.
The ``flags'' indicate the presence of other muons before or after the considered $\mu$-stop, signals from C3 counter (caused by muon decay), Ge-Ge coincidences, ``busy'' signals, etc., which are used in off-line analysis. The time interval between the $\mu$-stop and the Ge signal is used to distinguish between a Prompt (P) or a Delayed (D) signal, and to evaluate the muon lifetime in the target. In addition, each subdirectory includes a count rate of all HPGe detectors, counters C0-C3, number of $\mu$-stops, dead time and other slow-control information.

As mentioned above, isotope-enriched targets are strongly preferable. On the other hand, measurements with the natural targets are also important for comparison and identification of numerous $\gamma$-lines in all spectra measured. Table~\ref{tab:targets} lists all the targets used in the present study.

\begin{table}[h]
  \caption{Targets used in the present \mucap\ experiments. In the fourth column, the mass of the target element is presented (not the total mass, in the case of compound powders).}
  \label{tab:targets}
\begin{tabular}{c|c|c|c|c}
\hline
target                     &  enrichment &composition           &	element &thickness      \\
          &   &           &mass &mg/cm$^2$      \\\hline
$^{48}$Ti 	               &95.8\%       &TiO$_2$ powder	    & 1.0 g          &250  \\
$^{76}$Se 	               &92.4\%       &Se granules           & 5.0 g          &800 \\
$^{\rm nat}$Se             &--           &Se granules           & 5.0 g          &800 \\
$^{106}$Cd                 &63.0\%       &Cd metallic foil     	& 5.0 g	         &800 \\
$^{\rm nat}$Cd             &--           &Cd metallic foil      & 5.0 g          &800 \\
$^{82}$Kr 	               &99.8\%       &Kr gas        	    & 0.7 l (1 atm.) &26.1 \\
$^{\rm nat}$Kr             &--           &Kr gas                & 1.0 l (1 atm.) &37.3 \\
$^{150}$Sm                 &92.6\%       &Sm$_2$O$_3$ powder    & 2.0 g	         &320 \\
$^{\rm nat}$Sm             &--           &Sm$_2$O$_3$ powder    & 2.0 g	         &320 \\
\hline
\end{tabular}
\end{table}

\section{EXTRACTION OF RATES} \label{sec:extractionofrates}

The time evolution of $\gamma$ transitions in the daughter nucleus resulting from $\mu$-capture in the $(A,Z)$
target atom, is uniquely determined by the lifetime $\tau$ of the muonic atom. The total
muon disappearance rate $\lambda_{\rm tot}$ is
\beq\label{eq:lambdatotal}
\lambda _{\rm tot} = 1/\tau = \lambda _{\rm cap} + H\cdot\lambda _{\rm free},
\eeq
where $\lambda _{\rm cap}$ and $\lambda _{\rm free}$ are the total $\mu$-capture rate and free muon decay rate (0.4552$\times 10^6$ s$^{-1}$), and $H$ is the Huff factor~\cite{Huff-AnnPhys16-1961}, which is introduced to account
for the phase-space change in the decay of a muon bound in the atomic K-shell. A list of Huff factors used in the present analysis is given in Table~\ref{tab:Huff}.

\begin{table}[h]
  \caption{The values of Huff factor used in our analysis of the \mucap\ experiments taken from~\cite{Huff-AnnPhys16-1961}.}
  \label{tab:Huff}
\begin{tabular}{c|c}
\hline
target                     &  Huff factor   \\\hline
Ti 	               &0.98       \\
Se 	               &0.96          \\
Kr 	               &0.95  \\
Cd                 &0.92      \\
Sm                 &0.89     \\
\hline
\end{tabular}
\end{table}

\begin{figure}[h]
 \begin{center}
   \includegraphics [width= \linewidth]{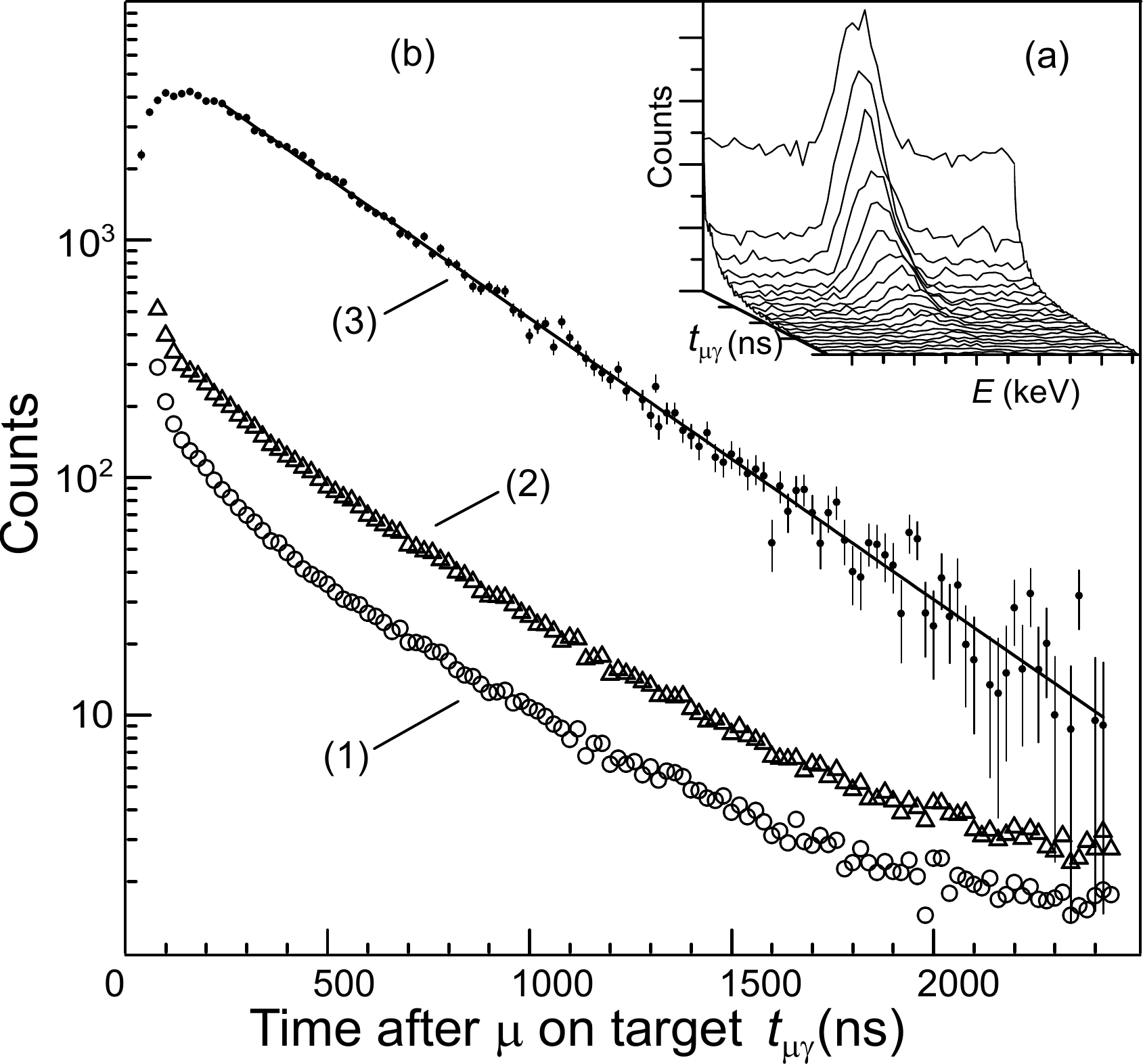}
 \caption{Time evolution of the spectrum fragment -- 227 keV $\gamma$-line following OMC in \ti48. (1) -- integral of central part of the fragment (line itself + background under it); (2) -- area of the background around the line and (3) -- integral of the $\gamma$-line fitted with five parameter Gaussian for each time slice.}
  \label{fig:time-evolution}
 \end{center}
\end{figure}

The procedure used to extract the total capture rate was developed in our previous work~\cite{Fynbo_NP,Egorov-CJP56-2006}. It is based on observing the time evolution of the most intense $\gamma$-line. An example of the procedure is shown in Fig.~\ref{fig:time-evolution}.

The measured \gray count is histogrammed in two dimensions: energy, and delay time $t_{\mu\gamma}$, which is divided into bins of a few nanoseconds (see Fig.~\ref{fig:time-evolution} inset). We then study a fragment of this histogram for the energy region containing a Line of Interest (LOI). The simplest way to analyze this is to plot the total count in the fragment containing the LOI (curve 1 in Fig.~\ref{fig:time-evolution}). However, this count contains not only \gray from muon capture, but also some background events. As the background may originate from different processes, it may therefore have a different time behavior (curve 2) from the signal; this means that curve 1 is not a pure exponential.

A more precise method involves fitting the energies in the fragment around the LOI to a Gaussian with five parameters:
\beq
f(E) = \frac{S_{\rm i}}{\sqrt{2\pi}\sigma}\cdot e^{\frac{-(E - E_{\rm 0})^2}{2{\sigma}^2}} + F_{\rm 0} + F_{\rm 1}\cdot(E - E_{\rm 0}),
\eeq
where $E_{\rm 0}$ and $\sigma$ represent parameters of the $\gamma$ peak, $F_{\rm 0}$ and $F_{\rm 1}$ are the background parameters and $S_{\rm i}$ is the extracted signal event count for energy fragment $i$.
Fitted $S_{\rm i}$ values are presented as (3) in Fig.~\ref{fig:time-evolution}. Their exponential behavior is much more evident.

It should be mentioned that, for small values of $t_{\mu\gamma}$, the exponential behavior is affected, not only by non-ideal detector timing, but also by the nonzero lifetime of the excited state (see, for instance, 183 keV $\gamma$ line in Fig.~\ref{fig:Matrix_76Se}). The later part of the time distribution is also slightly affected by random coincidences (which in our case are almost negligible). In order to estimate systematic error caused by the above effects and the analysis procedure, the exponent parameter $\tau$ was extracted several times: for each detector channel (Ge-1A...Ge-4B), for two different TDC and for different time bin intervals. An example of such extraction is given in Table~\ref{tab:fitresults}. All of these results were averaged. The uncertainty was taken to be whichever was greater: the standard deviation of the weighted average value, or that of the statistical mean.

\begin{table}[h]
  \caption{Part of the fit results of the 245 keV $\gamma$-line following OMC in $^{82}$Kr (in Ge-2A, 100 MHz TDC).}
  \label{tab:fitresults}
\begin{tabular}{c|c|c|rcr}
\hline
$\tau$                  & $\Delta\tau$ &$\chi^{2}$/\sl DoF     &\multicolumn{3}{c}{time bins} \\
\text{[ns]}                  &  [ns] &     &from &-- &to       \\\hline
143.92 	               &0.54       &2.04	    & 5 &-- &70         \\
141.84 	               &0.69       &1.79           & 5 &-- &65          \\
142.83             &0.95         &1.86           & 5 &-- &60          \\
143.73                 &1.09      &1.97     	& 10 &-- &60	         \\
142.28             &0.77         &1.91      & 10 &-- &65          \\
143.51 	               &0.66      &1.95        	    & 10 &-- &68 \\
\hline
\end{tabular}
\end{table}

All $\gamma$-lines observed for a given target should produce the same $\tau$ value (which is a property of the target, but not of the line). To obtain a central $\tau$ value for a target, individual measurements can be averaged as explained above. Some lines in the $\gamma$ spectrum have ambiguous origins: an alternative reaction (as an example, being origin from the $\mu$-capture in isotope $(A+1,Z+2)$ with one neutron emission) or unknown admixture ($\mu$-capture in other surrounding materials, the same energy of the $\gamma$-line produced by the different way ). In these cases, the $\tau$ value extracted will be unreliable, and should be excluded from the analysis.

As it was mentioned above in Eq.~\ref{eq:lambdatotal} muon disappearance rate $\lambda_{\rm tot}$ includes the capture rate $\lambda_{\rm cap}$ which in its turn consists of partial rates to the all states of the daughter nucleolus, including its ground state ($j$=0), excited bound states ($j$=1, 2, ...), as well as higher excited states decaying with a nucleon(s) emission:
\beq\label{eq:lambdacap}
\lambda _{\rm cap} = \lambda_{\rm cap}(0n) + \lambda_{\rm cap}(1n) + \lambda_{\rm cap}(2n) + \lambda_{\rm cap}(1p) + ...,
\eeq

For the theoretical NME calculations it is important to know the partial capture rate $\lambda_{j}$ or at-least the relative intensity $P_{j}$ of the $\mu$-capture to the particular daughter state $j$, i.e., a capture with no nucleon emission (the first term in Eq.~\ref{eq:lambdacap}).
Defining the percentage of the muons captured to the particular $j$ state
with respect to the total number of muons stopped in the given target isotope as a yield $Y_j$, the partial capture rate could be expressed as:
\beq
 \lambda_{j}  = {Y_{j}\cdot\lambda_{\rm tot}}\;\;,
\eeq
\noindent
and the relative probability
\beq\label{eq.probability}
P_{j}  = \frac{\lambda_j}{\lambda_{\rm cap}}\cdot 100\% = \frac{Y_{j}\;\lambda_{\rm tot}}{\lambda_{\rm cap}}\cdot 100\% .
\eeq

While the total muon disappearance rate $\lambda_{\rm tot}$ can easily be extracted using the above method, the extraction of partial $\mu$-capture
rates to the individual daughter states is less obvious,
and requires a special approach.
In a hypothetical ideal case the value of $Y^\mu_{j}$ could be found as the ratio:

$Y_j$ = $\frac{\rm Intensity\;of\;\gamma-rays\;discharging\;the\;state\;j}{\rule{0mm}{2.5mm} \rm Intensity\;of\;\mu X-rays\;=\;the\;number\;of\;stopped\;muons}$. \rule[-4mm]{0mm}{9mm}

In the reality, the $j$ state is populated not only by direct muon capture, but also by $\gamma$-feeding from the upper levels. That is why to extract the partial rates, it is necessary to evaluate in detail the relative intensities of the $\gamma$ transitions between the bound states of the daughter nucleus (see for instance~\cite{Gorringe-PRC60-1999,Siegbahn-1965,PDG-2012}). Unfortunately, the relevant level schemes of odd-odd final nuclei are poorly known\footnote{As normally they are not populated in conventional $\beta$-decay.}, and therefore, in the present study one is forced to restrict oneself to the low-energy and well-established part of the level schemes.

Finally, the population of the state by OMC is found as a difference between the total intensity of the $\gamma$ transitions discharging $I^\gamma_{\rm out}$ and feeding $I^\gamma_{\rm in}$ the particular level $j$
(here and thereafter all the intensities $I$ of $\gamma$-transitions, as well as the product yields $Y$ are normalized by the number of $\mu$-stops in the given chemical element):

\begin{equation}\label{eq.yields}
  \setlength{\unitlength}{1mm}
  \put(-40,-10){                         
  \begin{picture}(70,20)(0,-10)
  \put (19.0, 0.0){\makebox(0,0)[r]{\scriptsize\bf Excited state, \sl j}}
  {\linethickness{1pt}
  \put (21.0, 0.0){\line(1,0){38}}}
  \put (13.0,+2.0){\makebox(0,0)[lb]{\scriptsize\sl feeding: }}
  \put (8.0,-2.5){\makebox(0,0)[lt]{\scriptsize\sl de-excitation: }}
  \put (24.0, 6.0){\vector(0,-1){ 6}}
  \put (22.0, 6.0){\line(1,0){10}}
  \put (24.5,+3.0){\makebox(0,0)[l]{\scriptsize$\gamma_1$}}
  \put (29.0, 8.0){\vector(0,-1){ 8}}
  \put (29.5,+4.0){\makebox(0,0)[l]{\scriptsize$\gamma_2$}}
  \put (27.0, 8.0){\line(1,0){9}}
  \put (34.0,10.0){\vector(0,-1){10}}
  \put (34.5,+5.0){\makebox(0,0)[l]{\scriptsize$\gamma_3$}}
  \put (27.0, 0.0){\vector(0,-1){ 8}}
  \put (27.5,-4.0){\makebox(0,0)[l]{\scriptsize$\gamma_4$}}
  \put (25.0,-8.0){\line(1,0){10}}
  \put (32.0, 0.0){\vector(0,-1){ 6}}
  \put (32.5,-3.0){\makebox(0,0)[l]{\scriptsize$\gamma_5$}}
  \put (35.0,-6.0){\line(-1,0){6}}
  \put (49.0,+1.0){\makebox(0,0)[b]{\scriptsize$\sum(I^\gamma_i)=I^\gamma_{\rm in}$}}
  \put (49.0,-1.0){\makebox(0,0)[t]{\scriptsize$I^\gamma_{\rm out}=I^\gamma_{\rm in}+Y_{j}$}}
  \put (63.0, 0.0){\vector(-1,0){3}}
  \put (63.0, 0.0){\line(0,-1){5}}
  \put (64.0, 0.0){\makebox(0,0)[l]{\scriptsize $Y_{j}$}}
  \put (63.0,-6.0){\makebox(0,0)[t]{\scriptsize\sl OMC}}
  \end{picture}
}
\end{equation}

The intensity of each\footnote{It should be mentioned that in principle there could be unobserved transitions (not intensive, highly converted, hidden by other intensive lines, etc.), but we believe that this contribution is not crucial.} particular $\gamma$-line $I^\gamma_{i}$ normalized by the number of $\mu$-stops in the given isotope with the enrichment $\varepsilon$ is extracted from the detailed analysis of D-spectra as:

\beq\label{eq.intensityofgamma}
I^\gamma_{i}  = \frac{S^\gamma_{i}}{\eta_i\;\varepsilon{\sum\nolimits_{n}{I(K_n)}}}\;\;,
\eeq
\noindent
where $\eta_i$ is the relative detector efficiency, and  $\varepsilon{\sum\nolimits_{n}{I(K_n)} }$ is the sum of intensities of the \muXray\ $K$-series in the given isotope (cf. Fig.~\ref{fig:Se-prompt-delayed}), and $S^\gamma_{i}$ is an area of the $\gamma$ peak in the spectrum.
In case when the peak execution is doubtful (because of admixture from other isotopes/$\gamma$-lines, \muXray s, etc.) its intensity could be estimated from intensity of other lines genetically related with the given one, using the branching information from the level scheme in~\cite{nndc}.

As an example, Table~\ref{tab:partial-rates-48Ti} illustrates the detailed OMC data for the $^{48}$Ti target. The $\gamma$-lines which could not be identified and analyzed properly are not presented in that work.

The uncertainty of the partial capture rate was estimated as a full systematic error of indirect measurement (using the well known methods~\cite{PDG-2012,Hudson-CERN,Leo-TNPE}).
The total error is a combination of the gamma-peak intensity error, uncertainties on the intensities of the
K-series $\mu$X-rays (used for gamma-peak normalization), the dispersions of the weighted average values
of the $\lambda_{\rm tot}$ and $\lambda_{\rm cap}$, the efficiency determination error, and the
uncertainty on the relative intensity for the gamma-peak (taken from~\cite{nndc} database). The contribution of the uncertainty to the results in detailed balance is given in Table~\ref{tab:uncertainty}.
\begin{table*}[ht]
  \caption{Estimated uncertainty.}
  \label{tab:uncertainty}
\begin{tabular}{c|c|c|c}
\hline\hline
Extracted                & Origin of the  &Uncertainty, $\%$  &Comments \\
value                 & uncertainty &     &       \\\hline
$\Delta I^{\gamma}_i$ 	               &peak area       &1--25	    & depends on the line intensity and the background     \\
	               &detector efficiency       &5--20           & at low and high energy region it is increased         \\
            &relative branching $^\ddag$        &2--30           &  taken from~\cite{nndc}         \\\hline
$\Delta Y_{j}$                 & sum of the $\Delta I^{\gamma}_{\rm in}$ and $\Delta I^{\gamma}_{\rm out}$     &22--43    	& proportional to the total number of the feeding and discharging $\gamma$-lines 	  \\\hline
$\Delta P_{j}$            &$\Delta Y_{j}$        &22--43      &         \\
	               &$\Delta \lambda_{\rm tot}$     &0.06--0.6        	    & dominated by admixture of heavier isotope in the enriched target  \\
        &$\Delta \lambda_{\rm cap}$     &0.06--0.6              & dominated by admixture of heavier isotope in the enriched target \\
\hline\hline
\multicolumn{4}{l}{$^\ddag$) was used only in case when the execution of the $\gamma$-peak is doubtful.}\\
\end{tabular}
\end{table*}

The total capture rate of the produced nuclei $\lambda_{\rm cap}(xn)$ (as well as total capture rate to all bound states $\lambda_{\rm cap}(0n)$) and the probability $P_{\rm cap}$ that nuclei will be produced in $(\mu^-,\nu~xn)$
reactions are:
\begin{align}
\label{eq:lambdaXn}
\begin{split}
\lambda_{\rm cap}(xn)= {Y_{\rm cap}(xn)}\cdot{\lambda_{\rm tot}}\;,
\\
\lambda_{\rm cap}(0n) = \sum_{j=0,1,..}\lambda_j = {Y_{\rm cap}(0n)}\cdot{\lambda_{\rm tot}}\;,
\end{split}
\end{align}
\noindent
and
\beq
P_{\rm cap} = \frac{\lambda_{\rm cap}(xn)}{\lambda_{\rm cap}}\cdot 100\% = \frac{Y_{\rm cap}(xn)\;\lambda_{\rm tot}}{\lambda_{\rm cap}}\cdot 100\% .
\eeq

Unfortunately, the intensity of delayed $\gamma$-rays does not provide information on the partial capture rate to the ground ($\lambda_0$) states, as they are not followed by the delayed $\gamma$-rays.
Nevertheless, if the ground state of the daughter nuclei is unstable ($\beta^-$/$\beta^+$/EC active) the gamma radiation with intensity $I^{\gamma}$(U) following decay of long-lived states can be found in U-spectra. That allowing us to estimate the activity of the daughter nuclei and to extract their total yield $Y_{\rm cap}(xn)$:
\beq
Y_{\rm cap}(xn)  = {\frac{I^{\gamma}(U)}{BR(U)}}\cdot CF\;\;,
\eeq
\noindent
where $BR(U)$ is a branching ratio of the Uncorrelated $\gamma$-rays taken from the decay level scheme~\cite{nndc}.
Correction factor $CF$ depends on the ratio between the measurement time $t_{\rm exp}$ and the life-time of the ground state $\tau_{\rm g.s.}$. It takes into account the accumulation of long-lived activity and the resulting decay rates.
If the nucleus is relatively short-lived ($\tau_{\rm g.s.} \ll t_{\rm exp}$) -- then this factor is close to unity and could be neglected. Otherwise its estimation is more complicated (see for instance~\cite{Leo-TNPE}) and in addition should include the target activation during the tuning periods, deactivation at the beam interrupts and variation of the beam intensity.

The obtained results are presented in Table~\ref{tab:yields} (subsection~\ref{subsec:Se}). As with $\lambda_{\rm cap}$, values and errors of $\lambda_{\rm cap}(xn)$ (and therefore $P_{\rm cap}(xn)$) have been deduced as the weighted average values over all Uncorrelated lines.

\section{MEASUREMENT RESULTS}

\subsection{$^{48}$Ti target}

1.5 g  of $^{48}$TiO$_2$ powder material was packaged into a thin mylar bag and placed inside the target compartment (cf. Fig.~\ref{fig:target}). The titanium was composed of: $^{46}$Ti -- 0.8\%, $^{47}$Ti -- 1.2\%, $^{48}$Ti -- 95.8\%, $^{49}$Ti -- 1.4\%, $^{50}$Ti -- 0.8\%. Total exposure time was 73 hours, and the average rate of $\mu$-stops in the target was $12.8\cdot10^3\rm ~s^{-1}$.

\begin{figure}[h]
 \begin{center}
   \includegraphics [width= \linewidth]{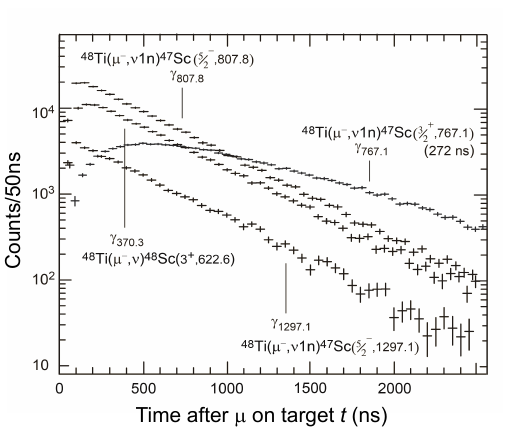}
 \caption{Time evolution of the intensities of some the strongest \gam-lines produced $\mu$-stop in \ti48. Note, the isomer of $^{47}$Sc (\jpi$=3/2^+$, 767.1~keV) decays with its own
 characteristic lifetime $\approx 272$~ns.}
 \label{fig:Ti-rates}
 \end{center}
\end{figure}

The $\mu$X-rays in the prompt P-spectra, and the delayed $\gamma$-rays (in the D-spectra) that follow muon capture (corresponding to the de-excitation of the bound states of $^{48}$Sc) were investigated. The $\tau$ value was extracted, using the most intense delayed $\gamma$-lines. The results of the fit are shown in Fig.\ref{fig:Ti-rates}. The exponential behavior of three lines, which share the same $\tau$ parameter, can be clearly seen. The $\gamma$-line at 370.3 keV the lowest energy of the three, which explains why the line demonstrates the highest deviation at short delay times
(as explained in section~\ref{subsec:approach}, Fig.~\ref{fig:time-resolution}). For the fourth \gam-line (at 767.1 keV), the de-excitation of the corresponding isomeric state of $^{47}$Sc has a significant lifetime, leading to a non-exponential time dependence for this line.

The obtained values of lifetime and total \mucap\ rate are presented in Table~\ref{tab:total-caprates}, whereas the partial capture rates of the $^{48}$Ti($\mu^-$,$\nu$)$^{48}$Sc reaction are listed in Table~\ref{tab:partial-rates-48Ti}.

\renewcommand{\arrayrulewidth}{.8pt}
\renewcommand{\doublerulesep}{.2pt}
\begin{table*}[ht]
\caption{Muon lifetimes and \mucap\ rates $\lambda_{\rm cap}$ in different targets. Lifetimes have been extracted using only the strongest \gam-lines. Errors are statistical, extracted from the lifetime fitting procedure.
 The  capture rates have been deduced from the weighted average values of the life-times indicated by the symbol $<>$. In  the cases of the natural targets (right part of table), the mother nuclei are identified by their unique position within the isotopic chain and by making the additional assumption that a possible 3n-channel contribution from an adjacent isotope  is negligible, or, as in $^{112}$Cd, that the 2n-channel from $^{113}$Cd is suppressed by low isotopic abundance. In a few cases for the enriched targets (\ti48, \se76\ and \cd106) the reaction channels may be slightly contaminated with a \mucap\ reaction from the adjacent isotope. In these cases an extra systematic error of order $1 - 2\%$ may have to be added to the extracted lifetime and rates quantities. The same applies to $^{112}$Cd. This is indicated by an asterisk in the various rows. Note, that the 767.1~keV transition de-excites the isomeric state of $^{47}$Sc with a half-life of 272(8)ns.}
\label{tab:total-caprates}
\begin{tabular}{c|c|c|c|c||c|c|c|c|c}
\hline\hline
Target&Daughter &$E^\gamma_{i}$&$\tau$&$<\lambda_{\rm cap}>$&Target &Daughter&$E^\gamma_{i}$&$\tau$&$<\lambda_{\rm cap}>$\\
 &isotope&$\text{[keV]}$&$\text{[ns]}$&$\text{[$10^6 $s$^{-1}$]}$&&isotope&$\text{[keV]}$&$\text{[ns]}$&$\text{[$10^6 $s$^{-1}$]}$\\
\hline

$^{48}$Ti &$^{48}$Sc &370.3	&363.8(26) & &$^{\rm nat}$Se &&&&  \\
          &$^{47}$Sc &807.8	&359.7(28) & &               &$^{76}$As &164.7 &163.5(20) &5.68(7)\\
          &          &1297.1&358.0(40) & &               &$^{77}$As &215.5 &165.9(19) &5.59(7)  \\
          &$^{47^{*}}$Sc &767.1	&358(10)[+272]&&$^{80}$Se&$^{79}$As &109.7 &185.5(27) &4.96(7)\\\cline{4-5}
 (*)         &          &   &$<>=$361.1(24)&2.323(15)&$^{82}$Se &$^{81}$As&336.0&208.2(68)&4.37(14) \\\cline{1-5}\cline{6-10}

$^{76}$Se &$^{75}$As &198.6	&148.4(7)  & &$^{\rm nat}$Cd &&&& \\
          &          &264.7	&148.4(5)  & &               &$^{109}$Ag&311.4 &92.2(26)  &10.43(31)\\
          &          &279.5	&148.6(5)  & &               &$^{110}$Ag&483.7 &95.0(70)     &10.11(75)  \\
          &$^{74}$As &183.0	&148.5(13) & &$^{112}$Cd     &$^{111}$Ag&376.7 &99.4(24)  & \\\cline{4-5}
 (*) &          &      &$<>=$148.48(10)&6.300(4)   &&          &391.3 &99.5(22)  & \\\cline{1-5}
$^{106}$Cd&$^{105}$Ag&346.8 &73.2(5)   & &   (*)            &          &      &$<>=$99.45(5)&9.600(5) \\\cline{9-10}
          &          &433.2	&72.4(8)   & &$^{114}$Cd     &$^{113}$Ag&270.8 &102.2(16) & \\\cline{4-5}
 (*)      &          &      &$<>=$72.97(36)&13.28(7) &   &          &366.8 &101.9(19) & \\\cline{1-5}

$^{82}$Kr &$^{82}$Br &244.8	&142.9(6)  & &               &          &      &$<>=$102.07(15)&9.380(14)\\\cline{9-10}
          &$^{81}$Br &276.0	&142.6(3)  & &$^{116}$Cd     &$^{115}$Ag&255.5 &107.7(18) &8.86(15) \\\cline{6-10}
          &          &649.8	&143.5(17) & &$^{\rm nat}$Kr &&&&  \\\cline{4-5}
          &          &      &$<>=$142.68(37)&6.576(17)  &&$^{84}$Br &408.2 &160.1(27) &5.81(10)\\\cline{1-5}

$^{150}$Sm&$^{149}$Pm&114.0 &82.1(6)   & &$^{86}$Kr      &$^{85}$Br &345.2 &173.5(26) &5.33(8)\\\cline{6-10}
          &          &188.4	&82.3(10)  & \\
          &          &198.6 &83.0(10)  & \\
          &          &208.0	&83.4(19)  & \\
          &          &211.2	&81.8(9)   & \\
          &          &287.9 &83.5(16)  & \\
          &$^{148}$Pm&219.8	&83.1(21)  & \\
          &          &233.0	&81.7(21)  & \\\cline{4-5}
          &          &      &$<>=$82.3(5)&11.75(7) \\
\hline\hline
\end{tabular}
\end{table*}

The experiment devoted to measuring \mucap\ in $^{48}$Ti was performed twice, in 2001 and 2002. A great effort was made to achieve stability, and to acquire sufficient statistics to investigate the U-spectra. Unfortunately, in both experiments, the beam intensity was unstable. Most of the products of \mucap\ in $^{48}$Ti are long-lived or stable; the shortest-lived isotope, $^{48}$Sc, has a lifetime of 43.67 hours, comparable with the total exposure. Investigation of the Uncorrelated spectra in this case was therefore very complicated, making it impossible to extract the total capture yield from the U-spectra with appropriate precision.

\begin{table*}[ht]
\caption{The detailed OMC data for the $^{48}$Ti target. The values of the first six columns are taken from~\cite{nndc}.
 The last five columns show the extracted values: experimental intensities of $\gamma$-rays $I^\gamma_{i}$ and their sums de-exciting ($I^\gamma_{\rm out}$) or feeding ($I^\gamma_{\rm in}$) the intermediate daughter state $j$ in the $^{48}$Ti($\mu^-$,$\nu$)$^{48}$Sc reaction, as well as partial yields Eq.~(\ref{eq.yields}) and relative probabilities Eq.~(\ref{eq.probability}).}
\label{tab:partial-rates-48Ti}
\begin{tabular}{c|c|c|c|c|c||c|c|c|c|c}
\hline\hline
\multicolumn{2}{c|}{Daughter OMC state}&\multicolumn{2}{c|}{De-exciting $\gamma$ transition}&\multicolumn{2}{c||}{Final state}&\multicolumn{2}{c|}{De-excitation $^\ddag$}&\multicolumn{1}{c|}{Feeding $^\ddag$}&\multicolumn{2}{c}{Partial OMC $^\ddag$}\\\hline
$E_{j}$ 	&$J^\pi$& $E^\gamma_{i}$	&$I^\gamma_{i} (rel)$ & $E_j$ &$J^\pi$ &$I^\gamma_{i}\times10^{3}$ &$I^\gamma_{\rm out}\times10^{3}$ &$I^\gamma_{\rm in}\times10^{3}$ &$Y_{j}\times10^{3}$ &$P_{j}$ \\
 $\text{[keV]}$    && $\text{[keV]}$ & $[ \% ]$ &     $\text{[keV]}$  &&&&&& $[ \% ]$    \\
\hline
0   	&$6^+$		 &\multicolumn{8}{c}{ground state}		   \\\cline{1-11}
130.9	&$5^+$	     &130.9(4)   &100.0       &0       &$6^+$  &*      &*  &*      &*  &*              \\\cline{1-11}
252.4	&$4^+$	     &121.4(4)   &100.0       &130.9   &$5^+$  &*      &*  &*      &* &*      \\\cline{1-11}
622.6	&$3^+$	     &370.3(5)   &100.0       &252.4   &$4^+$  &*      &*  &57.7(523) &*         &*      \\\cline{1-11}
1142.6	&$2^+$	     &519.9(2)   &100.0       &622.64  &$3^+$  &28.8(57) &28.8(57) &19.2(136) &9.6(71) &1.185(877)      \\\cline{1-11}
1401.6	&$2^-$	     &259.1(2)   &4.2(11)     &1142.6  &$2^+$  &0.90(27) &22.3(86) &13.1(96) &9.2(57)  &1.136(707)     \\
        &            &779.0(2)   &100.0(11)   &622.6   &$3^+$  &21.4(52) &&&& \\\cline{1-11}
1891.1	&$3^-$  	 &489.3(3)   &72.5(20)    &1401.7  &$2^-$  &1.10(19) &3.10(132) &2.20(75) &0.90(49) &0.11(6)\\
        &            &748.3(4)   &13.7(20)   &1142.6   &$2^+$  &0.20(4) &&&& \\
        &            &1268.3(6)   &11.8(20)   &622.6   &$3^+$  &0.20(5) &&&& \\
        &            &1638.8(3)   &100.0(20)   &252.4  &$4^+$  &1.60(34) &&&& \\\cline{1-11}
2190.5	&$3^+,1^+$   &1567.7(3)  &28.2(13)   &622.6    &$3^+$  &1.90(84) &8.80(486) &4.4(8) &4.40(256) & 0.55(32)\\
        &            &1938.1(3)   &100.0(13)   &252.4  &$4^+$  &6.90(227) &&&& \\    \cline{1-11}
2275.5	&$2^+$  	 &1132.8(3)   &100.0(15)  &1142.6  &$2^+$  &4.0(16) &5.7(34) &--     &5.70(337) &0.71(42)  \\
        &            &1652.9(3)   &42.9(15)   &622.6   &$3^+$  &1.7(7) &&&& \\\cline{1-11}
2517.3	&$1^+$	     &1374.7(3)   &100.0      &1142.6  &$2^+$  &4.1(18) &4.1(18) &0.0 &4.10(181) &0.52(23) \\\cline{1-11}
2640.1	&$1,2^-$	 &1238.4(3)      &100.0   &1401.7  &$2^-$  &8.5(26) &8.5(26) &--     &8.50(265) &1.056(328)\\\cline{1-11}
2670.3	&$1^-,2^-$ 	 &1268.5(3)   &100.0    &1401.7    &$2^-$  &1.5(5) &1.5(5) &0.0 &1.50(47) &0.19(6)\\\cline{1-11}
2729.0	&$(4^+,5^+)$ &2476.6(8)   &100.0    &252.4     &$4^+$  &1.5(6) &1.5(6) &--     &1.50(63) &0.19(8) \\\cline{1-11}
2783.3	&$2^+$  	 &892.0(5)    &100.0(17) &1891.1   &$3^-$  &2.20(75) &3.8(27) &--     &3.80(267) &0.47(33)  \\
        &            &1381.9(5)   &58(4)   &1401.7     &$2^-$  &1.30(55) &&&& \\
        &            &2160.4(8)   &12(7)   &622.6      &$3^+$  &0.30(14) &&&& \\\cline{1-11}
2980.8  &$1^+$       &1838.3(3)   &100     &1142.6     &$2^+$  &4.3(24) &4.3(24) &0.0 &4.30(235) &0.53(29)\\\cline{1-11}
3026.2	&$(2,3)$	 &835.6(5)       &100(7)  &2190.5  &$3^+$  &4.4(8) &9.5(58) &--     &9.5(58) &1.175(717)\\
        &            &1624.4(5)   &17(5)   &1401.7     &$2^-$  &0.70(25) &&&& \\
        &            &1883.8(6)   &22(5)   &1142.6     &$2^+$  &1.0(3) &&&& \\
        &            &2403.7(6)   &78(5)   &622.6      &$3^+$  &3.4(10) &&&& \\\cline{1-11}
3056.5	&$1^+$       &1913.9(3)   &100     &1142.6     &$2^+$  &3.6(19) &3.6(19) &0.0 &3.60(192) &0.45(24)   \\\cline{1-11}
3149.9	&$1^+$  	 &2007.3(3)   &100     &1142.6     &$2^+$  &1.1(6) &1.1(6) &--     &1.10(63) &0.14(8)      \\\cline{11-11}
        &&&&&&&&&                                                                 &$\sum=8.40(157)$  \\
\hline\hline
\multicolumn{11}{l}{$^\ddag$) the results of the present work.}\\
\multicolumn{11}{l}{*) the $\gamma$-line is distorted by other intense lines, and cannot be extracted.}\\
\multicolumn{11}{l}{--) no reference data available.}\\
\end{tabular}
\end{table*}

\subsection{$^{76}$Se and $^{nat}$Se targets}\label{subsec:Se}

The targets for these measurements were made from metallic selenium powder (cf. Table ~\ref{tab:targets}) in a thin plastic bag. The enriched target was composed of: $^{74}$Se -- 0.07\%, $^{76}$Se -- 92.4\%, $^{77}$Se -- 1.17\%, $^{78}$Se -- 2.28\%, $^{80}$Se -- 3.44\%, $^{82}$Se -- 0.64\%. A measurement with a natural selenium target was also performed. The number of muon stops in the experiment with the enriched target was about $1.7\cdot10^4\rm ~s^{-1}$ and the exposure time was $154\rm~h$. The number of stops with the natural selenium target was about $2.5\cdot10^4\rm ~s^{-1}$; total acquisition time for the spectra was $43\rm~h$.

\begin{figure}[h]
 \begin{center}
   \includegraphics [width= \linewidth]{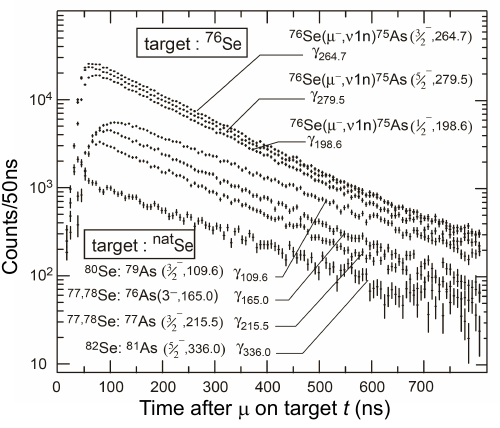}
 \caption{Evolution of $\gamma$-lines intensity with time after $\mu$-stop in Se targets. The figure shows two separate measurements, one with an enriched $^{76}$Se target (top graphs) and one with a natural Se target. In the latter the main components of the initial Se-isotopes are indicated. The identification is based on the different lifetimes  and the isotopic abundances. The deterioration of the time resolution below $\approx 200$~keV \gray\ energy is clearly observed.}
 \label{fig:Se-rates}
 \end{center}
\end{figure}

The $\gamma$-lines (corresponding to de-excitation of the bound states of $^{76}$As, as well as the most intensive $\gamma$-lines in the D-spectra of the enriched and natural targets) in P- and D-spectra were investigated. Measured \mucap\ rates in different Se isotopes are shown in Fig.~\ref{fig:Se-rates}.

The curves in the figure are divided into two parts: the $\gamma$-lines corresponding to the enriched target $^{76}$Se, and the $\gamma$-lines measured with the $^{nat}$Se target. The exponential behavior of the lines corresponding to the enriched target, which share the same $\tau$ parameter, is clearly seen. The $\gamma$-lines corresponding to other isotopes have a different slope, and also show deviation at short times, corresponding to charge collection in the low-energy region. Extracted life-times and \mucap\ rates $\lambda_{\rm cap}$ in different selenium nuclei are presented in Table~\ref{tab:total-caprates} and in~\cite{Zinatulina-BRAS72-2008}. Partial capture probabilities for the $^{76}$Se($\mu^-$,$\nu$)$^{76}$As reaction are presented in Table~\ref{tab:partial-rates-76Se}.

\begin{table*}[ht]
\caption{Partial OMC probabilities to different excited states of $^{76}$As. There are three separate tables side by side listing the excitation energies (columns one) and the \jpi\ values (columns two). Most of the excitation energies and \jpi\ values are taken~\cite{nndc}. The values in square brackets $[~]$ are from~\cite{Thies-PRC86-2012}. Partial capture probabilities $P_{j}$ (columns three) as percentages of the total capture rate $\lambda_{\rm cap}$ taken from Table \ref{tab:total-caprates}. The  partial capture probabilities integrated over the excitation region of $\approx 1$~MeV (bottom row) amount to $\approx 12\%$ of the total rate $\lambda_{\rm cap}$.}
\label{tab:partial-rates-76Se}
\begin{tabular}{c|c|c||c|c|c||c|c|c}
\hline\hline

$E_{j}$ 	&$J^\pi$&	$P_{j}$	&	 $E_{j}$ &$J^\pi$&		 $P_{j}$		&	  $E_{j}$	 &$J^\pi$&	$P_{j}$ \\
 $\text{[keV]}$    && $[ \% ]$ &     $\text{[keV]}$             && $[ \% ]$   &  $\text{[keV]}$ & &$[ \% ]$    \\
\hline
  0.0	&$2^-$	     &g.s.& 	     499.6	 &$[1^+,2^-]$&0.99(36)	&    802.4	 &$(1^-,2^-,3^+)$ &0.17(10)       \\
120.3	&$1^+$	     &0.32(12)&		 505.2	&$(2,3)^+$	     &0.25(6) &  863.3	 &$1^+$	         &0.27(20) 	   \\
122.2	&$(1)^-$     &0.21(11)	&	 517.6	&$(1,2^+$)	     &0.24(11)  &893.2	 &$(1^-,2^-,3^+)$ &0.23(10)         \\
165.0	&$(3)^-$	 &0.54(31)&      544.0	&$(2,3)^-$	     &0.39(24)&	 924.7	 &($\leq$ 3)$^-$	 &0.24(10)         \\
203.5	&$(0,1)^+$	 &0.08(4)	&    610.0	 &(1,2,$3^-$)	 &0.68(20) & 939.7	 &$(1,2,3)$	    &0.33(25)  		      \\
280.3	&$(1,2)^+$	 &0.11(5)	&  	 640.1	 &($1^-$,$2^-$)	 &0.18(9)&	 958.4&	 $\leq$ 3        &0.13(8)      \\
292.6	&$(2,3,4)^-$	 &0.05(1)&	 669.1	 &($1^+$,$2^+$)	 &0.64(20)&	 985.5	 &$(1,2,3)^+$	 &0.21(12)          \\
328.5	&$(3,4)^-$	 &0.09(4)	&    681.1	&$(1-4)$	     &0.33(10) & 1026.2	 &$[1^+,3^+]$	 &0.96(24)      \\
352.4	&$(3)^-$	 &0.05(2)&       734.4	&($\leq$ 4)$^-$  &0.08(4)  & 1034.2	 &$(1,2,3)^+$    &0.13(8)             \\
401.8	&$(1,2)^+$	 &0.41(26) &     751.8	   &($0^-$,1,2)	 &0.37(19)  &1064.5	 &$1^+$	        &0.23(15)         \\
436.8	&$(1,2,3)^-$ &0.28(13)&		 756.6	 &$(0^+,3^+)$	 &0.26(10)	&           &&              \\
447.2	&$(1,2)^+$	 &0.46(23) &     774.4	 &$[1^+,3^+]$	 &0.23(11) 	      &         &&              \\\cline{9-9}
471.0	&$(2)^-$	 &0.05(4)&       793.6	 &$(1,2,3)^+$	 &0.20(15)  &     &&          $\sum=11.99(105)$\\
\hline\hline
\end{tabular}
\end{table*}

As mentioned above, (cf. in~\ref{subsec:UPD} and in~\ref{sec:extractionofrates}) the intensities of $\gamma$ transitions in U-spectra could be used to extract the production rates of the ground and isomeric states for the
($\mu^-$ + $^{76}$Se) reaction. The results are presented in Table~\ref{tab:yields}.

\begin{table}[h]
\caption{The total capture rates to the excited states $\lambda_{\rm cap}(xn)$ and the probabilities $P_{\rm cap}$ of isotopes/isomers in the $\mu$-capture by $^{76}$Se and $^{150}$Sm. Column one: isotopes/isomers produced in $\mu$-capture reaction; column two: decay type of isotope/isomers; column three: lifetime of the isotopes/isomers; column four: total $\mu$-capture rates $\lambda_{\rm cap}(xn)$ of isotopes/isomers; column five: the probabilities $P_{\rm cap}$ as percentages of capture rate $\lambda_{\rm cap}$.}
\label{tab:yields}
\begin{tabular}{c|c|c|c|c}
\hline\hline

isotope         &decay type 	     &$T_{1/2}$	&  $\lambda_{\rm cap}(xn)[10^6$s$^{-1}]$ & $P_{\rm cap}$  \\\hline
$^{76}$As      & $\beta^{-}$	 &26.3~h  &0.86(3)& 13.65(255)\\
$^{75m}$As  &IT   	 &17.6~ms   &0.41(7)& 6.5(11)\\
$^{75}$As  &\multicolumn{2}{c|}{stable}	    &\multicolumn{2}{c}{unmeasured}\\
$^{74}$As      &$\beta^{-}$, EC 	 &17.8~d  &1.1(2) & 17.5(32) \\
$^{73}$As  &EC& 80.3~d	    &\multicolumn{2}{c}{unmeasured}\\
$^{72}$As      &$\beta^{+}$    &26~h    &0.15(3)& 2.4(5) \\
$^{71}$As      &$\beta^{+}$	 &65.3~h  &0.061(18)&0.96(28)\\
$^{75m}$Ge &IT    	 &48~s    &0.047(13)&0.75(21)   \\
$^{75}$Ge      &$\beta^{-}$	 &82.8~min&0.054(2)& 0.86(3)  \\
$^{71m}$Ge &IT       &20~ms   &0.020(3)& 0.32(5)  \\
$^{74}$Ga      &$\beta^{-}$	 &8.1~min &0.026(6)& 0.40(9)\\
$^{72}$Ga      &$\beta^{-}$    &14.1~h  &0.026(7)& 0.40(11)\\
\cline{5-5}
&&&&$\sum$=43.7(43)\\
\hline
$^{150}$Pm      &$\beta^{-}$	 &2.68~h  &1.45(11)& 12.3(9) \\
$^{149m}$Pm     &IT	     &35~$\mu$s&1.80(31)& 15.3(26) \\
$^{149}$Pm      &$\beta^{-}$	 &53.1~h  &2.93(60)& 24.9(51)\\
$^{148}$Pm      &$\beta^{-}$   &5.37~d  &0.77(26)& 6.6(22)  \\
$^{148m}$Pm     &IT        &41.3~d  &0.10(2) & 0.85(17) \\
$^{148m}$Pm     &$\beta^{-}$	     &41.3~d  &0.21(6) & 1.79(51) \\
$^{149}$Nd      &$\beta^{-}$	 &1.73~h  &0.78(35)& 6.6(29)  \\
$^{148}$Nd      &\multicolumn{2}{c|}{stable}  &\multicolumn{2}{c}{unmeasured}  \\
\cline{5-5}
&&&&$\sum$=68.3(69)\\
\hline\hline
\end{tabular}
\end{table}

In the case when the half-life of the decaying nucleus is comparable with the exposure time (154 h), the value of the yield is corrected for the decay. The extracted probability $P_{\rm cap}$ of the bound states of $^{76}$As is in a good agreement with the sum of corresponding partial capture probabilities presented in Table~\ref{tab:partial-rates-76Se}. According to~\cite{Measday-PR354-2001} it is expected that $^{75}$As is produced with the highest yield, since the most probable interaction emits a single neutron. The yields of other nuclei decrease with the number of nucleons emitted. Unfortunately, as $^{75}$As is a stable nucleus, it does not reveal itself in the U-spectrum. This also applies to the long-lived isotope $^{73}$As.

\subsection{$^{106}$Cd and $^{nat}$Cd targets}

Both cadmium targets were made of metallic foil (cf. Table ~\ref{tab:targets}). The enriched target was composed of: $^{106}$Cd -- 63\%, $^{108}$Cd -- 0.6\%, $^{110}$Cd -- 4.5\%, $^{111}$Cd -- 4.7\%, $^{112}$Cd -- 10.8\%, $^{113}$Cd -- 4.0\%, $^{114}$Cd -- 10.3\%, $^{116}$Cd -- 1.8\%. The fragment of the P-spectrum shown in Fig.~\ref{fig:isotope-shift} demonstrates the isotopic shift of $K\alpha_1$ and $K\alpha_2$ $\mu$X-ray doublet. The \cd106 enrichment of the target extracted from these spectra, 63(2)\%, agrees with the certificate data. The \cd106\ content of the enriched target is 50 times higher than its natural abundance (1.25\%).
The exposure with the enriched target was $101\rm~h$; the number of $\mu$ stops was $7.6\cdot10^3\rm ~s^{-1}$. With the natural target it was $45\rm~h$ and $9\cdot10^3\rm ~s^{-1}$, respectively.

\begin{figure}[h]
 \begin{center}
   \includegraphics [width= \linewidth]{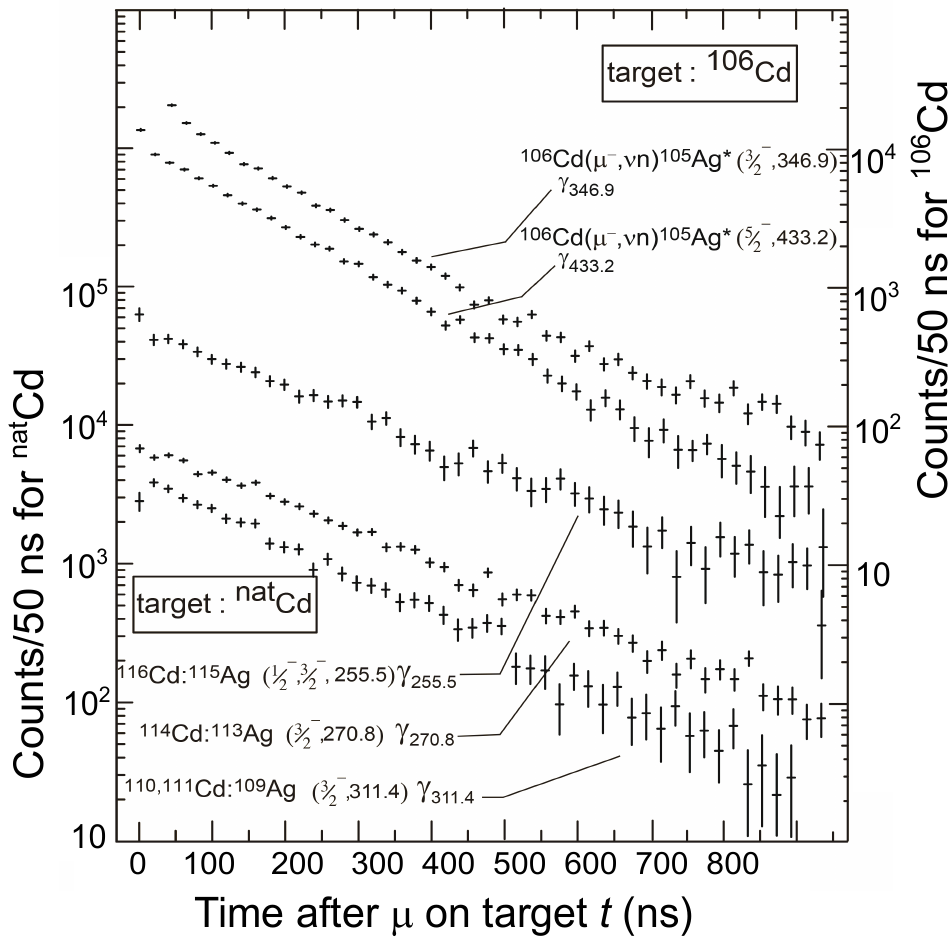}
 \caption{Evolution of $\gamma$-lines intensity with time after $\mu$-stop in Cd targets. The two upper plots correspond to the enriched target; the three lower plots, to the natural target. In the latter, the main components of the initial Cd isotopes are indicated. The identification is based on the different lifetimes and the isotopic abundances.}
 \label{fig:Cd-rates}
 \end{center}
\end{figure}

The evolution of the strongest $\gamma$-lines following OMC in \cd106 and $^{nat}$Cd is shown in Fig.~\ref{fig:Cd-rates}.
The difference in the exponential behavior (except for the two top lines, both of which correspond to $^{106}$Cd) is obvious, and is explained by the differing $\tau$ values for the different isotopes. The measured total \mucap\ rates for different cadmium isotopes are presented in Table \ref{tab:total-caprates} (see also~\cite{Zinatulina-BRAS72-2008}).
The muon lifetime in \cd106 was measured with the
enriched target, while the muon lifetimes for different
cadmium isotopes were measured with the natural target. The lifetime in $^{108}$Cd could not be measured because of its low abundance in the target, 0.89\%.

The extracted partial muon capture probabilities are listed in Table~\ref{tab:partial-rates-106Cd}.
Analysis of D-spectra showed many \gam-lines corresponding to the ROI. However, extracting the partial capture rates for this region was challenging, due to contamination from other \gam-lines and from \gray produced following OMC in the ($\mu^{-}$ + $^{106}$Cd) reaction. Such results are not presented in the Table.

\begin{table*}[ht]
\caption{Partial OMC probabilities to different excited states of $^{106}$Ag. There are three separate tables side by side listing the excitation energies (columns one) and the \jpi\ values (columns two), both taken from~\cite{nndc}. Partial capture probabilities $P_{j}$ (columns three) as percentages of the total capture $\lambda_{\rm cap}$ taken from Table \ref{tab:total-caprates}. The $^{106}$Ag has two states - $1^+$ g.s. and isomeric sate $6^+$ $^{106m}$Ag, so the  partial capture probabilities integrated over the excitation region of $1.3$~MeV (bottom row) amounts to $\approx 14\%$ of the total rate $\lambda_{\rm cap}$, but the sum of the partial capture probabilities corresponding to discharge only to the $1^+$ g.s. amounts to $\approx 12\%$ of the total rate $\lambda_{\rm cap}$.}
\label{tab:partial-rates-106Cd}
\begin{tabular}{c|c|c||c|c|c||c|c|c}
\hline\hline

$E_{j}$ 	&$J^\pi$&	$P_{j}$		& $E_{j}$ &$J^\pi$&		$P_{j}$	 & $E_{j}$	&$J^\pi$&	 $P_{j}$	 \\
 $\text{[keV]}$    && $[ \% ]$  &      $\text{[keV]}$             && $[ \% ]$    &  $\text{[keV]}$ && $[ \% ]$   \\
\hline
0   	&$1^+$		     &g.s.&	        542.6	&$(6)^+$	     &0.39(26)&      861.6	&$(3^-,4,5^-)$	 &0.31(13)       \\
89.7	&$6^+$		     &isom.state&             565.1	&$(2,3)^-$	     &1.07(66)&      917.4	&$1^-,2^-,3^-$	&0.30(17)   \\
110.7	&$(2)^+$	     &0.67(47)&     596.0	&$1^-,2^-,3^-$	 &0.60(49)  &	 936.5	&$(0^-,1^-)$    &0.23(13)  	    \\
205.9	&$(3)^+$         &0.37(34)&     597.3	&$1^-,2^-,3^-$	 &0.72(52)&      1003.4	&           	&0.34(20)   \\
234.6	&$(1^+,2^+,3^+)$ &0.64(41)&     602.8	&$(2^-,3^-)$	 &0.60(32)&	     1063.2	&$(4^+,5^+)$	&0.26(11)  	  \\
277.0	&$1^-,2^-$	     &1.43(115)&    661.3	&$1^-$ 	         &0.61(33)&	     1133.9	&$(4^+)$	    &0.08(6) \\
364.4	&$(2,3)^-$	     &0.51(41)&     698.2	&$1^-,2^-,3^-$	 &0.36(23)&  	 1145.1	&$(2^-,3^-,4^-)$ &0.18(12)         \\
389.1	&$3^+$	         &0.66(50)	&   730.4	&($2^+$)         &0.34(13) &  	 1303.3	&$(5^-,6^-,7^-)$ &0.26(15)     \\
416.6	&$(1^-)$	     &0.92(51)& 	741.6	&$(3^+,4^+)$	 &0.43(22)&      1329.5   &$(3^-,4^-)$    &0.37(20)  \\
425.0	&$(2,3)^-$	     &0.57(37)&  	765.2	&$6^-$	         &0.23(19)&       &&        \\\cline{9-9}
449.1	&$(4)^+$	     &0.22(10)&  	769.7	&$1^-,2^-,3^-$	 &0.30(17) &      && $\sum$(to~ $1^+$ ~g.s.)=12.36(201)  \\
468.8	&	             &0.06(4)&      812.0	&($3^-$)	    &0.21(18)&        && $\sum$(to~ $1^+$ and $6^+$)=14.26(220)    \\

\hline\hline
\end{tabular}
\end{table*}

The U-spectra in the ($\mu^-$ + $^{106}$Cd) reaction have been analysed. We do not observe the \gam-lines that would follow electron capture in the isomeric state of $^{106}$Ag.
Nevertheless, the (indirect) yield of this metastable state is non-zero - about 2\%, which follows from the balance of $\gamma$-rays in D-spectrum.
In U-spectra we observe the $\gamma$-lines following $\beta$-decay and EC of the ground states of $^{106}$Ag, $^{105}$Ag and $^{104}$Ag. All these lines, unfortunately, are contaminated with other strong $\gamma$-lines of very similar energy and therefore it is impossible to extract their intensity with appropriate precision.

\subsection{The Sm and Kr results}\label{subsec.SmKr}

For the measurements with Sm the targets ($^{nat}$Sm and enriched $^{150}$Sm) were packed in thin mylar films containing samarium oxide (cf. Table \ref{tab:targets}). The enriched $^{150}$Sm target was composed of: $^{144}$Sm -- 0.10\%, $^{147}$Sm -- 0.60\%, $^{148}$Sm -- 0.50\%, $^{149}$Sm -- 2.30\%, $^{150}$Sm -- 92.6\%, $^{152}$Sm -- 3.20\%, $^{154}$Sm -- 0.70\%. A test measurement with the $^{nat}$Sm target was used only to identify the $\gamma$-lines. Total irradiation of a $^{150}$Sm target was 68 hours with $\mu$ stops number of $4\cdot10^3\rm ~s^{-1}$.

\begin{figure}[h]
 \begin{center}
   \includegraphics [width=\linewidth]{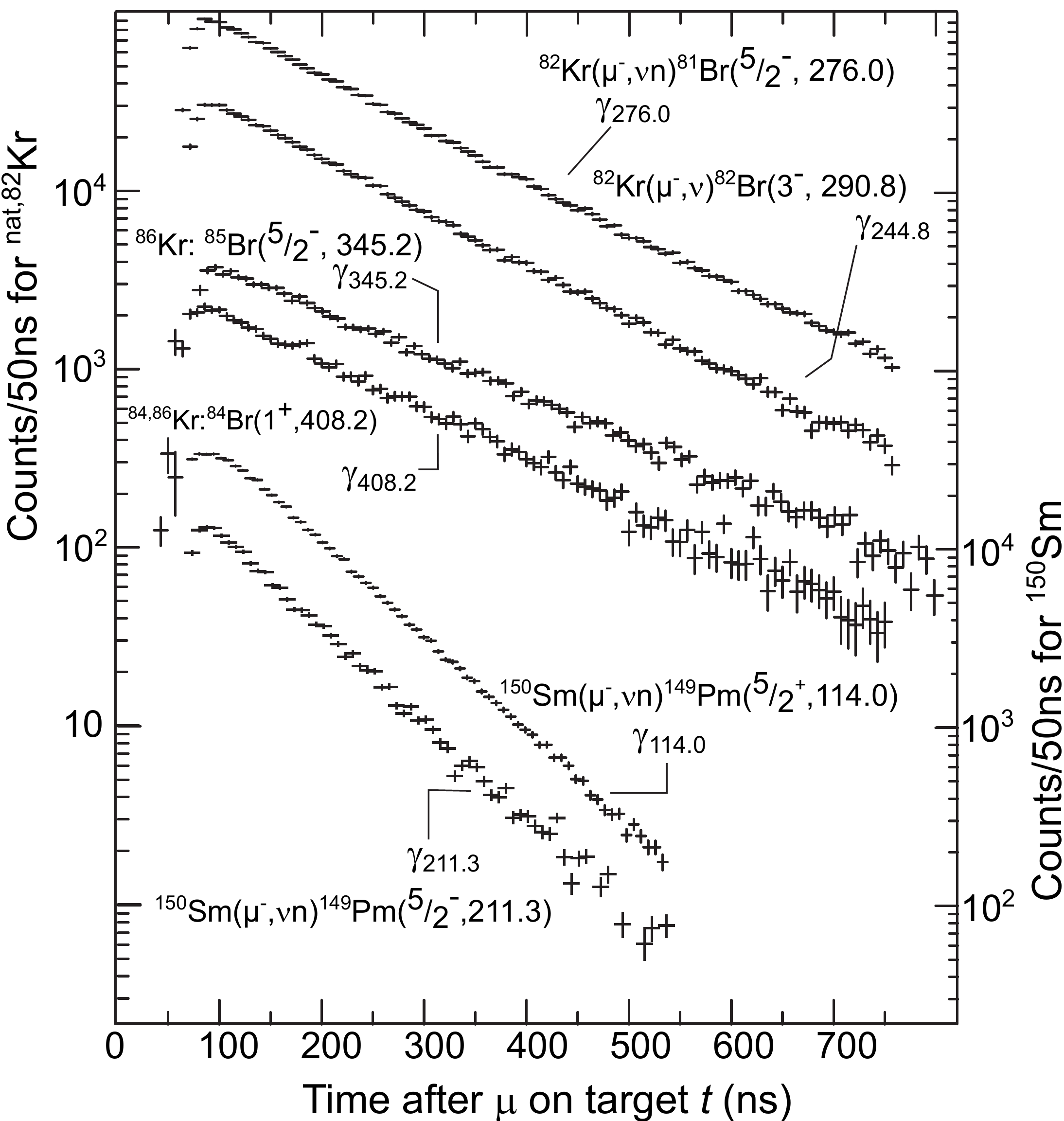}
 \caption{Measured \mucap\ rates on Kr isotopes and $^{150}$Sm. The measurements for $^{82}$Kr were performed with an isotopically
enriched material, while those for $^{84}$Kr and $^{86}$Kr used the natural composition of Kr gas. The
 total rates from the various Kr isotopes exhibit significant differences. The total rate on $^{150}$Sm (yield scale to the left) is about a factor of 2 larger than for the Kr cases. Note, $^{82}$Kr and $^{150}$Sm are the
 final nuclei in the $^{82}$Se and $^{150}$Nd \bb\ decays.}
 \label{fig:KrSm-rates}
 \end{center}
\end{figure}

The results of two strongest lines from D-spectra in $^{150}$Sm are shown in Fig.~\ref{fig:KrSm-rates}. Lifetime and \mucap\ rate $\lambda_{\rm cap}$ in $^{150}$Sm nucleus are presented in Table~\ref{tab:total-caprates}. The procedure of this experiment with $^{150}$Sm and $^{nat}$Sm was also presented in~\cite{Zinatulina-BRAS74-2010}.

As mentioned in~\cite{Zinatulina-BRAS74-2010}, it was impossible to extract partial capture rates, as no information about the excited $^{150}$Pm energy levels was available in~\cite{nndc} at the time of publication.
However,~\cite{Zinatulina-BRAS74-2010} includes all the delayed \gam\ transitions that we observed upon the irradiation of the $^{150}$Sm target by negative muons. Increased data on the energy levels has since become available~\cite{levelsPm150}, and will be used in our group's current studies.

Seven isotopes (isomers) in the $\mu$-capture process in $^{150}$Sm were identified in the U-spectra. The decay $\gamma$-ray intensities of these isotopes were used to determine the yields of these nuclei. The obtained isotope yields are listed in Table~\ref{tab:yields}.

The experiment with $^{nat}$Kr and enriched $^{82}$Kr requires special construction of a gas target, in order to prevent
any loss of this expensive noble gas; to minimize absorption of low-energy \gam-rays; and to ensure maximum percentage of muon stops in the gas, but not in the entrance window or
other construction materials~\cite{Zinatulina-AIP1572-2013,Zinatulina-AIP942-2007}.
The isotopic composition of the enriched target was the following: $^{78}$Kr -- 0.01\%, $^{80}$Kr -- 0.05\%, $^{82}$Kr -- 99.8\%, $^{83}$Kr -- 0.05\%, $^{84}$Kr -- 0.01\%, $^{86}$Kr -- 0.01\%.

For the measurements on $^{82}$Kr and $^{nat}$Kr, 0.7 liters and 1 liter of gas respectively were kept under a pressure of 1 atm. in a vessel, placed inside a similar setup to that used for the solid targets (cf. Fig.~\ref{fig:target}). The total exposure time with $^{82}$Kr was $122\rm~h$; for $^{nat}$Kr total exposure was $41\rm~h$. The number of $\mu$-stops was $1.5\cdot10^4\rm ~s^{-1}$ and $5\cdot10^3\rm ~s^{-1}$, respectively.

The total capture rates curves for different isotopes of Kr are shown in Fig.~\ref{fig:KrSm-rates} and numerical results are presented in Table \ref{tab:total-caprates}. Unfortunately, too many low-lying excited states of $^{82}$Br are populated in this reaction. Moreover, parallel reactions with neutron(s) emission produce $\gamma$-lines in the same ROI. As a result, we have not succeeded in extracting partial capture rates in the 0n channel with sufficient precision. Therefore, we present only muon lifetimes.

\section{CONCLUSIONS}

The experimental investigation of OMC in $^{48}$Ti, $^{76}$Se and $^{nat}$Se, $^{106}$Cd and $^{nat}$Cd, $^{82}$Kr and $^{nat}$Kr, $^{150}$Sm and $^{nat}$Sm was performed using the PSI muon beam and HPGe $\gamma$ detectors.

Time and energy distributions for $\mu$X- and $\gamma$-rays were measured precisely.

The time evolution of specific \gam-lines, as well as the intensities of Prompt $\mu$X-lines, and Delayed and Uncorrelated $\gamma$-lines, were analyzed. As a result, the total capture rates and partial capture probabilities were extracted for different isotopes, and the yield of radioactive daughter nuclei was estimated. The results are expected to be useful in calculating NME’s for \bb\ decay.

The extraction of the partial capture rates in the ($\mu^{-}$ + $^{150}$Sm) reaction is in progress.
It has been proposed that the $^{82}$Kr rates will be remeasured at the RCNP~\cite{Hashim-RCNP} negative muon beam (MuSIC, Osaka, Japan), with the MuSIC and UTM groups~\cite{Ejiri-PR338-2000,Hashim-PRC97-2018} in the near future.

Theoretical efforts are already in progress to incorporate the results presented here into calculations within the QRPA framework.

\section{AKNOWLEDGMENTS}

This work was supported in part by Russian Foundation for Basic Research (grant 06-02-16587).

The authors are grateful to H. Ejiri, V. Kuzmin, O. Civitarese and F. Simkovic, and appreciate their participation in the fruitful discussion of this work.

Authors are very grateful to Cheryl Patrick for the fruitful discussion of the paper and especially for the English proofreading.

\bibliography{muon-capture}

\end{document}